\documentclass[letterpaper,12pt]{article}
\usepackage{amsmath}
\usepackage{amssymb}
\usepackage{amsfonts}
\usepackage{setspace}
\usepackage{bbm}
\usepackage{color}
\usepackage{enumerate}
\usepackage{mathrsfs}
\usepackage[longnamesfirst]{natbib}
\usepackage[bottom]{footmisc}
\usepackage{graphicx}
\usepackage{subcaption}
\usepackage[hypertexnames = false]{hyperref} 
\usepackage[shortlabels]{enumitem}
\usepackage[usenames,dvipsnames,svgnames,table]{xcolor}
\usepackage{multirow}
\usepackage[margin=1.25in]{geometry}
\usepackage{tikz}
\usetikzlibrary{topaths,calc}
\usepackage{enumitem}
\usepackage{subcaption}

\usepackage{tikz}  
\usetikzlibrary{matrix,shapes,decorations.pathreplacing,fit,backgrounds}

\newlist{steps}{enumerate}{1}
\setlist[steps, 1]{label = Step \arabic*:}

\newtheorem{lemma}{Lemma}[section]
\newtheorem{proposition}{Proposition}[section]

\newtheorem{remark}{Remark}[section]
\newtheorem{example}{Example}[section]
\newenvironment{proof}[1][Proof]{\noindent \textbf{#1.} }{\  \rule{0.5em}{0.5em}}
\newcommand{\mb}[1]{\mathbb{#1}}

\newcommand{\mr}[1]{\mathrm{#1}}

\newcommand{\mc}[1]{\mathcal{#1}}

\newcommand{\ul}[1]{\underline{#1}}
\newcommand{\ol}[1]{\overline{#1}}

\hypersetup{
     colorlinks   = true,
     citecolor    = Blue,
     urlcolor     = Black,
     linkcolor    = Blue
}

\makeatletter
\renewcommand\paragraph{\@startsection{paragraph}{4}{\z@}%
                                    {0pt \@plus1ex \@minus.2ex}%
                                    {-1em}%
                                    {\normalfont\normalsize\bfseries}}
\makeatother

\usepackage{float}

\newcommand{\E}{\mathbb{E}}
\newcommand{\R}{\mathbb{R}}
\graphicspath{{pic/}}
\DeclareGraphicsExtensions{.pdf,.jpg,.png}
\newcommand{\bpm}{\begin{pmatrix}}
\newcommand{\epm}{\end{pmatrix}}

\usepackage{bibunits}
\usepackage[longnamesfirst]{natbib}
\shortcites{Bloom1997}

\definecolor{InvisibleRed}{rgb}{0.97, 0.92, 0.92}
\definecolor{InvisibleGreen}{rgb}{0.92, 0.97, 0.92}
\definecolor{InvisibleBlue}{rgb}{0.90, 0.92, 0.97}

\definecolor{MediumRed}{rgb}{0.925, 0.345, 0.345}
\definecolor{MediumGreen}{rgb}{0.32, 0.6, 0.5}
\definecolor{MediumBlue}{rgb}{0, 0.125, 0.376}

\definecolor{DarkBlue}{rgb}{0, 0.125, 0.376}
\definecolor{DarkRed}{rgb}{0.376, 0.125, 0}

\begin{document}

\defaultbibliography{evtc}
\defaultbibliographystyle{chicago}
\begin{bibunit}

\title{%
Externally Valid Policy Choice\thanks{%
This paper was first submitted to conferences and circulated for comments on April 20, 2022. We are grateful to the Co-Editor, three anonymous referees, Xiaohong Chen, Lars Hansen, Toru Kitagawa, Oliver Linton, Konrad Menzel, Francesca Molinari, Evan Munro, Jonathan Niles-Weed, Tom Sargent, Martin Schneider, J\"org Stoye, Andrei Zeleneev, and seminar and conference audiences at Bristol, Cambridge, Chicago, Cornell, Kantorovich Initiative, NAWM2023, NYU, Queen Mary, Queensland, and Stanford for helpful comments and suggestions. Support from the National Science Foundation via grant SES-1919034 is gratefully acknowledged.}}

\author{%
{Christopher Adjaho\thanks{%
 Department of Economics, New York University. \texttt{ca2384@nyu.edu}}
 \quad
 Timothy Christensen\thanks{%
 Department of Economics, Yale University. \texttt{timothy.christensen@yale.edu}}
 }
}

\date{November 25, 2026}

\maketitle

\begin{abstract}  
\singlespacing
\noindent
We consider the problem of estimating personalized treatment policies that are \emph{externally valid} or \emph{generalizable}: they perform well in target populations that differ from the experimental (or training) population from which the data are sampled. We first show that welfare-maximizing policies for the experimental population are robust to a certain class of shifts in the distribution of potential outcomes between the experimental and target populations (holding characteristics fixed). We then develop methods for estimating policies that are robust to shifts in the joint distribution of outcomes and characteristics. In doing so, we highlight how treatment effect heterogeneity within the experimental population shapes external validity. 

\medskip 


\end{abstract}

\pagenumbering{arabic}

\newpage

\section{Introduction}

There is increasing interest across many disciplines in personalized treatment policies.\footnote{In economics, the literature goes back to \cite{Manski2004} and includes \cite{BhattacharyaDupas}, \cite{Stoye2012}, \cite{KitagawaTetenov2018}, \cite{MTM}, and \cite{AtheyWager2021}. References in other fields include \cite{QianMurphy2011}, \cite{ZZRK}, \cite{SwaminathanJoachims}, and \cite{Kallus2017,Kallus2021}, among others.} The typical objective is to estimate a policy mapping individual characteristics into a treatment choice to maximize overall welfare. The assumption underlying this literature is that the \emph{target} population in which the policy is to be implemented is the same as the \emph{experimental} (or training) population from which the data are sampled.\footnote{We refer to the distribution from which the data are sampled as ``experimental'', with the understanding that the analyst could in fact be using observational (i.e., non-experimental) data.} While many policy learning algorithms have good welfare guarantees when the target and experimental populations are identical, the resulting policies can perform poorly if the target population differs from the experimental population. 

There are several reasons why the target and experimental populations may differ, echoing well-known concerns about the external validity of experiments \cite[see, e.g.,][]{BannerjeeDuflo2009,Deaton2010,Allcott2015}. For instance, data may be collected under experimental conditions that differ from real-world settings where the policy is to be implemented. Additionally, delays between data collection and implementation can result in distribution shifts. Experiments may also use easily measurable variables (e.g., test scores) to define outcomes, whereas the policymaker may be concerned with more difficult to quantify outcomes (e.g., overall academic achievement). Furthermore, experiments may be run in selected sub-populations that are not representative of the broader target population.

We consider the problem of learning personalized treatment policies that are \emph{externally valid} or \emph{generalizable}: they perform well in target populations different from the experimental population from which the data are sampled. We allow for shifts in both the distribution of potential outcomes and characteristics between the experimental and target populations. We propose methods for estimating externally valid policies using experimental or observational data (where treatment may be endogenous). Beyond policy learning, our methods can be used as a stress test to evaluate the robustness of treatment policies to distribution shifts. Our findings also shed light on the important roles that treatment effect heterogeneity within the experimental population plays in shaping the generalizability of policies.

More formally, consider a policy $\tau$ mapping individual characteristics $X$ into a binary outcome $\tau(X)$, where $\tau(X) = 1$ indicates that treatment is to be assigned to an individual with characteristics $X$ and $\tau(X) = 0$ indicates otherwise.\footnote{We do not consider randomized policies where $\tau(X)$ can also take intermediate values in $(0,1)$.} Following \cite{Manski2004}, policies are typically evaluated using a \emph{social welfare} criterion
\begin{equation}\label{eq:social.welfare.criterion}
 \mr{W}(\tau;P) = \E_P[ Y_1 \tau(X) + Y_0 (1-\tau(X))] \,,
\end{equation}
where $Y_0$ and $Y_1$ denote the individual's untreated and treated potential outcomes, and $\E_P[\,\cdot\,]$ denotes expectation under the distribution $P$ of $(X,Y_0,Y_1)$ in the experimental population.\footnote{We implicitly assume throughout that $\E_P[|Y_0|], \E_P[|Y_1|] < \infty$, so that $\mathrm{W}(\tau;P)$ is well defined.} It is also sometimes of interest to evaluate policies by their \emph{welfare gain} relative to a policy in which no one is treated:
\begin{align}
 \mr{WG}(\tau;P) & = \E_P[Y_1 \tau(X) + Y_0 (1-\tau(X))] - \E_P[Y_0] \notag \\
 & = \E_P[ (Y_1 - Y_0) \tau(X)]. \label{eq:welfare.gain.criterion}
\end{align}
The typical objective is to learn a policy that maximizes (\ref{eq:social.welfare.criterion}) or (\ref{eq:welfare.gain.criterion}) over a class $\mc T$  that may incorporate functional-form, budget, fairness, or other constraints. This literature relies on the assumption that the experimental and target populations are identical.

Our objective is to derive generalizable policies that deliver welfare guarantees over target populations that may differ from the experimental population. To this end, we replace criteria (\ref{eq:social.welfare.criterion}) and (\ref{eq:welfare.gain.criterion}) with the \emph{robust welfare} criterion
\begin{equation}\label{eq:robust.welfare.criterion}
 \mr{RW}(\tau;P) =  \inf_{Q \in \mc Q} \mr{W}(\tau;Q) \,,
\end{equation}
and the \emph{robust welfare gain} criterion
\begin{equation}\label{eq:robust.welfare.gain.criterion}
 \mr{RWG}(\tau;P) =  \inf_{Q \in \mc Q} \mr{WG}(\tau;Q) \,,
\end{equation}
where $\mc Q = \mc Q(P)$ is a set of target populations that are ``close'' to $P$ in a sense we make precise below.
We propose methods to learn robust policies $\tau_{\text{\tiny Robust}}$ that maximize $\mr{RW}$ or $\mr{RWG}$ over $\mc T$. This max-min approach can be viewed as a two-player zero-sum game in which the policymaker first chooses a policy and an adversary (``nature'') then chooses the target population $Q \in \mathcal Q$ to minimize welfare. By construction, then, $\tau_{\text{\tiny Robust}}$ delivers welfare guarantees uniformly over all target populations $Q \in \mc Q$.

A novel aspect of our approach is to define $\mc Q$ using Wasserstein distance. This has several advantages. First, the size parameter $\varepsilon$ used to define $\mc Q$ is the maximum difference in the average treatment effect (ATE) between the experimental population $P$ and target populations $Q \in \mc Q$, as we show  in Section~\ref{sec:neighborhood}. This makes $\varepsilon$ very interpretable and easily calibrated by the analyst. 
Wasserstein neighborhoods also lead to tractable characterizations of the robust welfare objectives (\ref{eq:robust.welfare.criterion}) and  (\ref{eq:robust.welfare.gain.criterion}). 

Section~\ref{sec:po} allows the conditional distribution of outcomes given characteristics to shift between the experimental and target populations, while the distribution of characteristics remains fixed. Propositions~\ref{prop:po.unbounded} and \ref{prop:po.gain.unbounded} characterize the robust criteria (\ref{eq:robust.welfare.criterion}) and (\ref{eq:robust.welfare.gain.criterion}) in this context. These characterizations imply that any policy that maximizes the usual welfare criteria (\ref{eq:social.welfare.criterion}) or (\ref{eq:welfare.gain.criterion}) is robust to a certain class of shifts in the distribution of outcomes. Hence, policy learning methods with good statistical guarantees under criteria (\ref{eq:social.welfare.criterion}) or (\ref{eq:welfare.gain.criterion}) \citep[see, e.g.,][]{Manski2004,QianMurphy2011,KitagawaTetenov2018,AtheyWager2021,MTM} also enjoy good guarantees under the corresponding robust criteria (\ref{eq:robust.welfare.criterion}) or (\ref{eq:robust.welfare.gain.criterion}). Appendix~\ref{sec:shift.x.known} extends these findings to settings with a known shift in characteristics. 

Section~\ref{sec:po.x} studies shifts in both outcomes and characteristics between the experimental and target populations. Proposition~\ref{prop:po.x.unbounded}  characterizes the robust criteria~(\ref{eq:robust.welfare.criterion}) and~(\ref{eq:robust.welfare.gain.criterion}) in this setting. In the worst case, nature moves individuals who benefit from treatment to the non-treatment side and moves individuals with negative treatment effects to the treatment side, prioritizing those close to the treatment/non-treatment frontier whose treatment effects are large in magnitude. This has two key implications for external validity. First, the location of individuals relative to the frontier matters. Robust policies tend to place individuals whose CATEs are small in magnitude close to the frontier and those whose CATEs are large in magnitude further away. Second, unobserved heterogeneity in treatment effects also matters. Because nature can shift individuals across the frontier, greater heterogeneity in treatment effects among individuals with the same characteristics gives nature more scope to lower welfare. 
By contrast, the usual criteria (\ref{eq:social.welfare.criterion}) and (\ref{eq:welfare.gain.criterion}) are invariant to the proximity of individuals to the frontier and to unobserved heterogeneity in treatment effects.  
An important take-away of this section is to be mindful of treatment effect heterogeneity when recommending policies in scenarios where distribution shifts are possible.

Section~\ref{sec:po.x.empirical} discusses empirical implementation. Here the robust criteria depend on the joint distribution of characteristics and individual treatment effects in the experimental population. This distribution is not identified without further assumptions. We first derive sharp bounds on the robust welfare criterion in the absence of identifying assumptions. We then introduce nonparametric estimators of the bounds and establish consistency and convergence rates.  Further, we provide a non-exhaustive set of methods for estimating robust policies under different identifying assumptions from the literature on distributional treatment effects (see \cite{AbbringHeckman2007} and references therein), allowing for experimental and observational data. We also provide convergence rates and regret guarantees for the estimated robust policies. 

Finally, Section~\ref{sec:extensions} considers two variants of our framework. First, Section~\ref{sec:nohet} explores an alternative neighborhood construction in which we shut down the ability of nature to shift distributions in a way that exploits unobserved heterogeneity in treatment effects. Second, most of our paper operates under the conventional assumption that the decision maker implements the (robust) welfare-maximizing policy irrespective of its cost. Section~\ref{sec:cost} discusses extensions to settings with a constant per-unit treatment cost. For both variants, we characterize robust welfare and discuss the estimation of robust policies.

We conclude in Section~\ref{sec:empirical} with an application that revisits the experiment of \cite{MKP2021}, which examined the effect of business training on rural Kenyan firms led by female entrepreneurs. Outcomes were measured over a four-year window following treatment, introducing a significant time lag between the experiment and any potential policy implementation. Our empirical findings corroborate some of our theoretical results about the role of heterogeneity in shaping generalizability. We also illustrate how the methods we develop can be used as a ``stress test'' for assessing the fragility or robustness of policies estimated using standard methods (e.g., empirical welfare maximization) to distribution shifts. For instance, researchers could report plots similar to those we present in Section~\ref{sec:empirical}, which compare the robust welfare of policies across families of target populations of different sizes.

\medskip

\paragraph{Related Literature.}

\cite{MoQiLiu} and \cite{Spini2021} study robustness to shifts in the distribution of $X$ over $f$-divergence neighborhoods. A key and arguably restrictive assumption underlying their approach is that the conditional distribution of outcomes given characteristics does not change between the experimental and target populations. By contrast, we allow the distribution of outcomes to shift as well.

Like us, \cite{Kido2022}, \cite{SZZB}, and \cite{QiPangLiu2022} seek robustness with respect to shifts in both outcomes and characteristics. These works and ours use different robustness sets which lead to different insights and different rankings of policies. \cite{Kido2022} restricts the conditional distribution of outcomes given $X$ to Wasserstein neighborhoods (pointwise in $X$), and restricts the marginal distribution of $X$ to $f$-divergence  (e.g., Kullback--Leibler divergence) neighborhoods. \cite{SZZB} constrain the joint distribution of outcomes and $X$ to Kullback--Leibler neighborhoods. \cite{QiPangLiu2022} constrain the ratio of the densities in the experimental and target populations. These constructions all prohibit extrapolation beyond the support of the experimental population, which may be restrictive in applications. The notion of neighborhood size is also more difficult to interpret for these constructions, whereas it has a clear interpretation in our setting. Section~\ref{sec:KL} discusses in more detail how Wasserstein distances address these shortcomings.
 Moreover, none of these prior works shed light on how treatment effect heterogeneity  affects generalizability, which is arguably the most important take-away from our work.

Moreover, \cite{Kido2022} considers shifts in outcomes only as well as shifts in both outcomes and $X$, as do we. For the former, his work and ours produce the same ranking if outcomes are unbounded from below. For the latter, \cite{Kido2022} constructs neighborhoods sequentially (first outcomes shift conditional on $X$, then $X$ shifts) whereas we construct them  jointly, allowing outcomes and $X$ to shift together. As a consequence, his robust criterion is invariant to unobserved heterogeneity in treatment effects whereas heterogeneity plays an important role in our characterization. A further interesting consequence is that robust policies in his setting can be learned using the methods of \cite{MoQiLiu}, which account only for shifts in $X$.

\cite{LSW2023} assume study participants are a non-random sample from the target population and model sampling bias,\footnote{See also \cite{Stoye2012} in the context of aggregate (i.e., not individualized) treatment policies.} i.e., how participants select into the sample. Their robustness set $\mc Q$ consists of unknown target distributions that are consistent with the experimental population given selection. We instead do not take a stand on the cause of the shift between $P$ and $Q$, allowing for a general class of unstructured shifts in distributions which could be caused by many mechanisms. In our approach, all the researcher needs to commit to is the maximum shift in the ATE between $P$ and $Q$ to be considered. Conversely, if the researcher has a good reason for committing to a specific mechanism causing the shift (e.g., sampling bias), then sharper results could be derived by tailoring the robustness set $\mathcal Q$ accordingly.

\cite{Munro2023} considers individualized policies in environments where agents strategically report their characteristics. His analysis attributes the distribution shift to strategic reporting by agents, which does not give rise to a distributionally robust problem. While we consider classes of deterministic policies, optimal policies with strategic agents may involve randomization.

Finally, our analysis in Section~\ref{sec:po.x.empirical} is related to, but conceptually distinct from, a literature on policy learning under ambiguity \citep[see, e.g.,][]{KZ2021,dadamo2022orthogonalpolicylearningambiguity,benmichael2025safepolicylearningextrapolation}, in which welfare is partially identified due to unobserved confounding or lack of overlap. In our setting, partial identification instead arises because robust welfare, the criterion we use to address external validity, depends on the joint distribution of potential outcomes in the experimental population.

\section{Wasserstein Neighborhoods}\label{sec:neighborhood}

In this section, we describe the set of target populations we work with and show that the neighborhood size has a clear interpretation as the maximal change in the ATE between the experimental and target populations. We then compare our approach with recent approaches using $f$-divergences (such as Kullback--Leibler divergence).

\subsection{Definition}

Let $Z :=  (X,Y_0,Y_1)$ take values in $\mc Z$ and let $d : \mc Z \times \mc Z \to \mb R_+ \cup \{+\infty\}$ be a metric on $\mc Z$. The \emph{Wasserstein distance} of order $p$ between $P$ and $Q$ is
\[
 d_{W,p}(P,Q) = \inf_{\pi \in \Pi(P,Q)} \E_\pi[d(Z,\tilde Z)^p]^{\frac{1}{p}}, \quad (1 \leq p < \infty),
\]
where $\Pi(P,Q)$ denotes all joint distributions for $(Z,\tilde Z)$ with marginals $P$ for $Z$ and $Q$ for $\tilde Z$. 
We will focus on the metric with $p=1$, so we drop the $p$ subscript and write $d_W(P,Q)$ in what follows. This is mainly to simplify presentation: Appendix~\ref{ax:extensions} presents generalizations to $p>1$. 

We define Wasserstein neighborhoods using 
\begin{equation}\label{eq:metric.general}
 d((x,y_0,y_1),(\tilde x, \tilde y_0, \tilde y_1)) = |y_0 - \tilde y_0| + |y_1 - \tilde y_1| + b(x,\tilde x),
\end{equation}
for some metric $b$. We use different choices of $b$ to handle robustness to shifts in outcomes only (as in Section~\ref{sec:po}) and shifts in outcomes and characteristics (as in Section~\ref{sec:po.x}). 
We define neighborhoods as
\begin{equation}\label{eq:wasserstein.neighborhood}
 \mathcal Q = \{ Q : d_W(P, Q) \leq \varepsilon\},
\end{equation}
where $\varepsilon > 0$ is a measure of neighborhood size. As we now show, the parameter $\varepsilon$ is the maximum change in the ATE as $Q$ varies over $\mc Q$. This interpretation holds irrespective of whether we seek robustness with respect to shifts in outcomes only or in both outcomes and characteristics. 

\subsection{Interpretation of Neighborhood Size}

First, we introduce some notation. Let $\mathcal Y$ denote the support of the potential outcomes.\footnote{By ``support'' we mean the set of all values the potential outcomes could conceivably take, as distinct from the measure-theoretic notion of support.} To simplify exposition, we present results for \emph{unbounded} potential outcomes for which $\inf \mathcal Y = -\infty$ and $\sup \mathcal Y = +\infty$. See Appendix~\ref{ax:bounded} for the case where at least one of $\inf \mathcal Y$ or $\sup \mathcal Y$ is finite, which is relevant, for example, for binary outcomes or non-negative outcomes such as earnings. Let $\Delta = Y_1 - Y_0$ denote the individual treatment effect.

\medskip

\begin{proposition}\label{prop:ate.unbounded}
Suppose that $\mc Q$ is defined using $d_W(P,Q)$ induced by (\ref{eq:metric.general}). Then
\begin{equation} \label{eq:bounds.ate}
 \begin{aligned}
 \inf_{Q \in \mc Q} \E_Q\left[ \Delta \right] 
 & = 
 \E_P \left[ \Delta  \right]  - \varepsilon, \\
 \sup_{Q \in \mc Q} \E_Q\left[ \Delta \right] 
 & = 
 \E_P \left[ \Delta \right]  + \varepsilon.
 \end{aligned}
\end{equation}
\end{proposition}

\medskip

\begin{remark}\normalfont
The bounds (\ref{eq:bounds.ate}) are sharp: there exist distributions $Q \in \mc Q$ for which $\E_Q\left[ \Delta \right] = \E_P\left[ \Delta \right] \pm \varepsilon$.  As such, Proposition~\ref{prop:ate.unbounded} provides a formal sense in which the neighborhood size $\varepsilon$ is precisely the maximum that the ATE can vary between the experimental population $P$ and target populations $Q \in \mc Q$. Proposition~\ref{prop:ate} in Appendix~\ref{ax:bounded} generalizes this interpretation to bounded outcomes as well.
\end{remark}

\begin{remark}\label{rmk:pretreat}\normalfont
A similar argument shows that $\varepsilon$ can also be interpreted as the maximum that the average untreated outcomes differ between the experimental and target populations:
\[
 \begin{aligned}
 \inf_{Q \in \mc Q} \E_Q\left[ Y_0 \right] 
 & = 
 \E_P \left[ Y_0  \right]  - \varepsilon, & & & 
 \sup_{Q \in \mc Q} \E_Q\left[ Y_0 \right] 
 & = 
 \E_P \left[ Y_0  \right]  + \varepsilon. 
 \end{aligned}
\]
\cite{Gechter2024} points out that in many circumstances one has information on the average untreated outcome in the target population. If so, one could calibrate $\varepsilon$ using the difference between the mean untreated outcomes in the experimental and target populations. Alternatively, one could calibrate $\varepsilon$ using the bounds on the ATE in the target population that \cite{Gechter2024} derives from information on untreated outcomes in the target population. 
\end{remark}

To give a quantitative example, consider the empirical application in Section~\ref{sec:empirical}. We are interested in the effect of a business training program on firm profits, which were measured over a four-year window following treatment. Therefore, any policymaker deciding whether to implement such a training program would be relying on evidence from an experiment conducted (at least) four years ago. This raises the possibility that the experimental and target populations can differ due to changes in the profit distribution over time. To quantify the drift, note that average profits for untreated firms one and three years after treatment are approximately 1,467 and 1,493 Kenyan Shillings (KSh), a difference of approximately 13 KSh per year. Remark~\ref{rmk:pretreat} suggests $\varepsilon \approx 50$ could be reasonable for a policymaker deciding whether to implement a policy four years after the experiment. In view of Proposition~\ref{prop:ate.unbounded}, $\varepsilon \approx 50$ represents slightly less than one third of the treatment effect on profits one year out of 157 KSh.

\subsection{Wasserstein or KL?}\label{sec:KL}

We close this section by contrasting our approach based on Wasserstein distance with the more familiar $f$-divergence (e.g., Kullback--Leibler (KL) divergence) used in a number of recent works.\footnote{Specifically, \cite{SZZB} constrain the joint distribution of $(X,Y_0,Y_1)$ using KL neighborhoods; \cite{MoQiLiu} and \cite{Spini2021} both assume there is no shift in the conditional distribution of $(Y_0,Y_1)|X$, but allow the marginal distribution of $X$ to vary over $f$-divergence neighborhoods; and \cite{Kido2022} allows the distribution of $(Y_0,Y_1)|X$ to vary over Wasserstein neighborhoods and constrains the distribution of $X$ using KL divergence.}

Consider an experiment in which subjects are assigned to a conditional cash transfer (CCT) program aimed at improving children's education outcomes. \cite{GS2017} provide a meta-analysis of CCTs. Suppose $X$ is households' distance from school and $(Y_0, Y_1)$ are educational outcomes without and with receiving cash transfers. Let $P$ denote the joint distribution over $(X, Y_0, Y_1)$ for an experimental population in a middle-income country, $Q$ the joint distribution in a target population that is another middle-income country, and $Q'$ the joint distribution in a target population that is a low-income country. 

\begin{table}[t]
\centering
\begin{tabular}{c|cccc}
$(x,y_0,y_1)$ & $P$ & $Q$ & $Q'$ & $P'$ \\ \hline \\[-10pt]
$(1, 50,50)$ & 0.05 & 0.40 & 0.05 & 0.1 \\
$(1, 50,80)$ & 0.05 & 0.40 & 0.05 & 0.1 \\
$(2, 50,50)$ & 0.40 & 0.05 & 0.05 & 0.4 \\
$(2, 50,80)$ & 0.40 & 0.05 & 0.05 & 0.4 \\
$(10, 50,50)$ & 0.05 & 0.05 & 0.40 & 0.0 \\
$(10, 50,80)$ & 0.05 & 0.05 & 0.40 & 0.0 \\
\end{tabular}
\caption{Illustrative example.}\label{tab:cct}
\end{table}

Table~\ref{tab:cct} presents a numerical example. In middle-income populations $P$ and $Q$, most students are 1 or 2 miles from their school, whereas most students in the low-income population $Q'$ are 10 miles from their school. All that matters for KL divergence is the relative probabilities over the support of the reference measure, not the actual values taken. Thus, both $Q$ and $Q'$ are the same KL-divergence from $P$:\footnote{Although we are considering shifts in the joint distribution of $(X,Y_0,Y_1)$, we have kept the values of $(Y_0,Y_1)$ fixed. This way we are effectively measuring shifts in the marginal distribution of $X$, so this critique is relevant for the constructions in \cite{MoQiLiu} and \cite{Spini2021}.}
\[
 KL(Q \|P) = KL (Q' \| P) \approx 1.456.
\]
However, it seems reasonable that low-income populations where 80\% of students live 10 miles from school are inherently ``further'' from middle-income populations in which 90\% of students live within 1 or 2 miles of their school. The fact that KL divergence does not reflect this difference limits its interpretability. However, this difference is reflected in Wasserstein distance:
\[
 \begin{aligned}
 d_W(P,Q) & = 0.70 , \\
 d_W(P, Q') & = 5.60 ,
 \end{aligned}
\]
where we have used the metric $b(x,\tilde x) = |x - \tilde x|$ in (\ref{eq:metric.general}).

Moreover, in the scenario where now the middle-income experimental population is given by $P'$, so that no student in the experimental population lives 10 miles from school, the distributions $Q$ and $Q'$ (and for that matter $P$) are not absolutely continuous with respect to $P'$. Correspondingly, we have
\[
 KL(Q\|P') = KL(Q'\|P') = KL(P\|P') = +\infty.
\]
In other words, a robustness analysis using KL divergence centered at $P'$ would never guard against shifts to $Q$ and $Q'$ (or $P$), since these are not within any ball of finite KL-divergence around $P'$. Our approach based on Wasserstein distance would guard against such shifts, because the distance between these distributions and $P'$ remains well defined, with 
\[
 \begin{aligned}
 d_W(P',Q) & = 1.40 , \\
 d_W(P', Q') & = 6.50 , \\
 d_W(P', P) & = 0.90.
 \end{aligned}
\]
While we have focused the above discussion on KL-divergences, similar critiques apply to other $f$-divergences.

\section{Shifts in Outcomes}\label{sec:po}

In this section we consider robustness with respect to shifts in the distribution of potential outcomes, holding characteristics fixed. This is relevant in a number of scenarios. First, it is relevant for designing policies when welfare is defined using outcomes that may differ slightly from those that are measured in the experimental population. For instance, the policymaker may care about overall educational attainment, while the experiment measures outcomes using a specific test score. It is also relevant when experimental conditions differ from the real-world setting in which the policy is to be implemented.\footnote{In these scenarios, we could think about individuals being characterized by $(X,Y_0,Y_1,Y_0^*,Y_1^*)$, where $(Y_0,Y_1)$ are the experimental-condition outcomes (or measured outcomes) and $(Y_0^*,Y_1^*)$ are the real-world setting outcomes (or policy-relevant outcomes). The distribution $P$ is the marginal for $(X,Y_0,Y_1)$, while $Q$ is the marginal for $(X,Y_0^*,Y_1^*)$.} 
Two extensions are presented in the appendix: see Appendix~\ref{ax:bounded} for bounded outcomes and Appendix~\ref{sec:shift.x.known} for known shifts in characteristics. 

We begin by introducing an appropriate neighborhood construction to handle this case. 
To allow for shifts in the distribution of potential outcomes conditional on characteristics while holding the distribution of characteristics fixed, we use Wasserstein distance induced by the metric
\begin{equation}\label{eq:metric.po}
 d((x,y_0,y_1),(\tilde x, \tilde y_0, \tilde y_1)) = |y_0 - \tilde y_0| + |y_1 - \tilde y_1| + \infty \times \mb I[x \neq \tilde x],
\end{equation}
where we adopt the convention that $0 \times \pm \infty = 0$.
This metric combines the $\ell^1$ norm to penalize the shifts in the distribution of outcomes conditional on characteristics, and a discrete metric on the support $\mc X$ of $X$ to prohibit shifts in characteristics.

The following result characterizes the robust welfare criterion.

\begin{proposition}\label{prop:po.unbounded}
Suppose that $\mc Q$ is defined using the Wasserstein distance $d_W(P,Q)$ induced by (\ref{eq:metric.po}). Then for any policy $\tau$, 
\[
 \mr{RW}(\tau;P) = \mr{W}(\tau; P)  - \varepsilon.
\]
\end{proposition}

An analogous result holds for robust welfare gain.

\begin{proposition}\label{prop:po.gain.unbounded}
Suppose that $\mc Q$ is defined using the Wasserstein distance $d_W(P,Q)$ induced by (\ref{eq:metric.po}). Then for any policy $\tau$ for which $P(\tau(X) = 1) > 0$,
\[
 \mr{RWG}(\tau;P) = \mr{WG}(\tau; P)  - \varepsilon,
\]
and $\mr{RWG}(\tau;P) = \mr{WG}(\tau;P) = 0$ otherwise.
\end{proposition}

We now discuss several implications of these results.

\begin{remark}\label{rmk:prop.po} \normalfont
Proposition~\ref{prop:po.unbounded} implies that any policy that maximizes the usual welfare criterion (\ref{eq:social.welfare.criterion}) must also maximize its robust counterpart (\ref{eq:robust.welfare.criterion}), irrespective of $\varepsilon$.  Moreover, the regret of any estimated policy $\hat \tau$ under criterion (\ref{eq:robust.welfare.criterion}) is equal to its regret under criterion (\ref{eq:social.welfare.criterion}):
\[
 \sup_{\tau \in \mc T} \mr{RW}(\tau;P) - \mr{RW}(\hat \tau;P) = \sup_{\tau \in \mc T} \mr W(\tau;P) - \mr W(\hat \tau;P), \quad \mbox{for all $\varepsilon > 0$.}
\]
Proposition~\ref{prop:po.gain.unbounded} provides an analogous result for welfare gain for policies $\tau$ for which $P(\tau(X) = 1) > 0$.
Hence, policy learning methods with good (statistical) regret guarantees under criteria (\ref{eq:social.welfare.criterion}) or (\ref{eq:welfare.gain.criterion}), such as those of \cite{Manski2004}, \cite{QianMurphy2011}, \cite{KitagawaTetenov2018}, \cite{AtheyWager2021}, and \cite{MTM}, to name a few, also enjoy good regret guarantees under their robust counterparts (\ref{eq:robust.welfare.criterion}) and (\ref{eq:robust.welfare.gain.criterion}). Propositions~\ref{prop:po} and~\ref{prop:po.gain} present versions of these results allowing for bounded potential outcomes. 
\end{remark}

For intuition for this result when $\mc Y = \R$, a ``worst-case'' distribution $Q \in \mathcal Q$ for which $\mathrm{W}(\tau;Q) = \mr{RW}(\tau;P)$ is the image measure of $P$ under the map
\begin{equation}\label{eq:t.map}
 T(x,y_0,y_1) = \begin{cases}
 (x, y_0 - \varepsilon, y_1), & \mbox{if $\tau(x) = 0$}, \\
 (x, y_0, y_1 - \varepsilon), & \mbox{if $\tau(x) = 1$}.
 \end{cases} 
\end{equation}
That is, the distribution of $X$ is held fixed, but $Y_0$ is reduced by $\varepsilon$ for untreated individuals and $Y_1$ is reduced by $\varepsilon$ for treated individuals. The net effect is to simply reduce welfare by $\varepsilon$, which is why the ordering over policies is the same for the usual welfare criterion and its robust counterpart. 
Interestingly, this is also a worst-case distribution in the related problem, where $\mc Q$ is defined as the set of distributions $Q$ for which the marginal distribution of $X$ is the same as under $P$ and, for each $x$, the conditional distribution for $(Y_0,Y_1)|X = x$ under $Q$ is constrained to a Wasserstein neighborhood of radius $\varepsilon$ of the conditional distribution under $P$, as in \cite{Kido2022}.

\section{Shifts in Outcomes and Characteristics}\label{sec:po.x}

We now turn to the problem of external validity when we allow for shifts in outcomes and characteristics. We begin in Section~\ref{sec:po.x.robust.criterion} by deriving the robust  criteria, and then explore their implications in Section~\ref{sec:po.x.implications}. In particular, we show how the distribution of CATEs and unobserved heterogeneity (within the experimental population) play distinct but important roles in shaping external validity.

Section~\ref{sec:po.x.empirical} discusses empirical implementation, allowing for experimental or observational data. Here the robust welfare criterion depends on the joint distribution of characteristics and individual treatment effects in the experimental population. As this distribution is not identified without further assumptions, we first derive sharp bounds on robust welfare without imposing identifying assumptions and show how the bounds can be estimated nonparametrically. We then provide approaches for estimating robust welfare under different identifying assumptions and establish convergence rates and regret guarantees for the estimated policies. As before, all results in this section are presented for the case in which potential outcomes are unbounded (i.e., $\inf \mathcal Y = -\infty$ and $\sup \mathcal Y = +\infty$ with $\mathcal Y$ the support of $Y_0$ and $Y_1$). See Appendix~\ref{ax:bounded} for the case where at least one of $\inf \mathcal Y$ or $\sup \mathcal Y$ is finite.

\subsection{Robust Criteria}\label{sec:po.x.robust.criterion}

To allow for shifts in the distribution of potential outcomes and characteristics, we use Wasserstein distance induced by the metric
\begin{equation}\label{eq:metric.po.x}
 d((x,y_0,y_1),(\tilde x, \tilde y_0, \tilde y_1)) = |y_0 - \tilde y_0| + |y_1 - \tilde y_1| + \|x - \tilde x\|,
\end{equation}
where $\|\cdot\|$ is a norm on $\mc X$. We say that $\tau \equiv 0$ if $\tau(x) = 0$ for all $x \in \mathcal X$, and $\tau \not \equiv 0$ if there is some $x \in \mathcal X$ for which $\tau(x) = 1$.

\medskip

\begin{proposition}\label{prop:po.x.unbounded}
Suppose that $\mc Q$ is defined using $d_W(P,Q)$ induced by (\ref{eq:metric.po.x}) and that $\E_P[\|X\|]$ is finite. Then for any policy $\tau$, 
\begin{equation} \label{eq:robust.po.x}
 \mr{RW}(\tau;P) =  \sup_{\eta \geq 1} \E_P \left[ \min \left\{ Y_0  + \eta h_0(X;\tau) , Y_1 + \eta h_1(X;\tau) \right\}   \right] - \eta \varepsilon ,
\end{equation}
where
\[
\begin{aligned}
 h_0(x;\tau) & =  \inf_{\tilde x \in \mc X: \tau(\tilde x) = 0} \|x - \tilde x\| \,, & & & 
 h_1(x;\tau) & =  \inf_{\tilde x \in \mc X: \tau(\tilde x) = 1} \|x - \tilde x\| \,,
\end{aligned}
\]
with the understanding that $h_0(x;\tau) = +\infty$ or $h_1(x;\tau) = +\infty$ if the infimum runs over an empty set. Moreover, for any $\tau \not \equiv 0$,
\begin{equation} \label{eq:equivalence}
  \mr{RWG}(\tau;P) = \mr{RW}(\tau;P) - \E_P[Y_0]\,,
\end{equation}
and $\mr{RWG}(\tau;P) = \mr{WG}(\tau;P) = 0$ if $\tau \equiv 0$.
\end{proposition}

The intuition for this criterion follows from the interpretation of Lagrangian duality as a two-player zero-sum game. The adversary (nature) can shift the distribution of $(X,Y_0,Y_1)$ at a price $\eta$ per unit of Wasserstein distance. Nature has a budget of $\varepsilon$ and two ways to spend it to lower welfare. The first option is to move treated individuals with $Y_1 - Y_0 > 0$ into the non-treatment region at a cost of $\eta h_0(X;\tau)$, and move untreated individuals with $Y_1  - Y_0 < 0$ into the treatment region at a cost of $\eta h_1(X;\tau)$. This choice is reflected by the minimum inside the expectation. Nature's second option is to shift outcomes at a cost of $\eta$ per unit of Wasserstein distance. Shifting outcomes is strictly profitable only when $\eta < 1$, which is why $\eta \geq 1$ in~\eqref{eq:robust.po.x}. If no individual is close enough to the frontier to be worth moving at $\eta = 1$, then nature spends its entire budget on shifting outcomes, and $\mathrm{RW}(\tau;P) = \mathrm{W}(\tau;P) - \varepsilon$ as in Proposition~\ref{prop:po.unbounded}. The supremum over $\eta$ enforces nature's budget constraint. Also note by~(\ref{eq:equivalence}) that both $\mathrm{RW}$ and $\mathrm{RWG}$ induce the same ordering over policies $\tau \not \equiv 0$.

The functions $h_0(x;\tau)$ and $h_1(x;\tau)$ represent the ``distance to non-treatment'' and ``distance to treatment'' under policy $\tau$ for an individual with characteristics $X = x$. Different policies $\tau$ induce different distances $h_0$ and $h_1$. We illustrate how to compute these for three popular classes of policies. To simplify exposition, we take $\mc X = \mb R^d$ and let $\|\cdot\|$ be the Euclidean norm.

\begin{example}[Linear Eligibility Score Policies] \label{ex:1} \normalfont
Consider a policy of the form
\[
 \tau(x) = \mb I[ \beta_0 + x'\beta_1 \geq 0],
\]
parameterized by $\beta_0$ (a scalar) and $\beta_1$ (a vector), where at least one element of $\beta_1$ is non-zero. The interpretation is that treatment is assigned when the eligibility score $x'\beta_1$ exceeds the threshold $-\beta_0$. 
Here $h_0(x;\tau)$ and $h_1(x;\tau)$ are the minimum distances from $x$ to the half-spaces $\{\tilde x : \beta_0 + \tilde x'\beta_1 \leq 0\}$ and $\{\tilde x : \beta_0 + \tilde x'\beta_1 \geq 0\}$, namely
\[
 h_0(x;\tau) = \frac{( \beta_0 + x' \beta_1)_+}{\|\beta_1\|}, \quad \quad h_1(x;\tau) = \frac{( \beta_0 + x' \beta_1)_-}{\|\beta_1\|},
\]
where $(a)_- = -\min\{a,0\}$ and $(a)_+ = \max\{a,0\}$. \hfill $\square$
\end{example}

\begin{example}[Threshold Policies] \normalfont
Consider a policy that assigns treatment when the values of certain characteristics are above or below given thresholds.
For illustrative purposes we consider a policy that depends on the first two components of $X$:
\[
 \tau(x) = \mb I[  s_1 (x_1 - \beta_1) \geq 0 \mbox{ and } s_2 (x_2 - \beta_2) \geq 0 ],
\]
for $s_1 = -1$ and $s_2 = 1$ and $\beta_1,\beta_2 \in \mb R$. Here $\tau(x) = 1$ if and only if $x_1 - \beta_1 \leq 0$ and $x_2 - \beta_2 \geq 0$. Similarly, $\tau(x) = 0$ if and only if $x_1 - \beta_1 > 0$ or $x_2 - \beta_2 < 0$. Hence,
\begin{align*}
 h_0(x;\tau) & = \min\left\{ (x_1 - \beta_1)_- \, ,  (x_2 - \beta_2)_+ \right\} \,, \\
 h_1(x;\tau) & = \sqrt{ (x_1 - \beta_1)_+^2 + (x_2 - \beta_2)_-^2 } \,. \tag*{$\square$}
\end{align*}
\end{example}

\begin{example}[Decision Trees] \label{ex:3} \normalfont
Decision trees are defined by recursive partitions of $\mc X$ into regions for which the values of characteristics lie above or below given thresholds. Treatment is then assigned depending on whether characteristics lie in certain subsets of the resulting partition. Policies based on decision trees may be expressed as
\[
 \tau(x) = \mb I[x \in \cup_{k=1}^K C_k],
\]
where each $C_k$ is a hyperrectangle. Different policies $\tau$ correspond to different sets $\cup_{k=1}^K C_k$. The non-treatment region $\mc X \setminus (\cup_{k=1}^K C_k)$ is itself the union of finitely many nonempty hyperrectangles, say $\cup_{l=1}^L \tilde C_l$. The squared distance from $x$ to a hyperrectangle $C$ is
\[
 \mr{dist}(x,C)^2 = \min_y \|x - y\|^2 \quad \mbox{subject to} \quad l_{i} \leq y_i \leq u_{i}, \quad i = 1,\ldots,d\,,
\]
where $l_{i}$ and $u_{i}$, $i = 1,\ldots,d$, define the boundaries of $C$ (we may have $l_{i} = -\infty$ or $u_{i} = +\infty$ for some $i$). Here $h_0(x;\tau)$ and $h_1(x;\tau)$ are then the minimum such distances from $x$ to the non-treatment and treatment hyperrectangles:
\[
 h_0(x;\tau) = \min_{1 \leq l \leq L} \mr{dist}(x,\tilde C_l)\,, \quad \quad h_1(x;\tau) = \min_{1 \leq k \leq K} \mr{dist}(x,C_k)\,. \tag*{$\square$}
\]
\end{example}

\bigskip

\bigskip

\noindent
We have so far been vague about the choice of norm $\|\,\cdot\|$ on $\mathcal X$. If the covariate vector $X$ contains variables whose units and scale are comparable to the outcomes (such as in our empirical application, where both are profits), our recommendation is to simply use the Euclidean norm. Otherwise, if $X$ contains variables whose units or scale are not comparable to outcomes, our recommendation is to use the weighted Euclidean norm
\[
 \|x - \tilde x\| = \sqrt{ (x - \tilde x)'\Sigma^{-1} (x - \tilde x) },
\] 
where
\[
 \Sigma = \frac{\mathrm{Cov}(X)}{\mathrm{Var}(Y_0)}.
\]
This weighting serves two purposes. First, it measures changes in $X$ on the same scale and in the same units as untreated outcomes. Second, it is invariant to rescaling or, more generally, invertible linear transformations of the $X$ variables. 

\begin{remark} \normalfont
By looking at the largest $\varepsilon$ for which $\mr{RW}(\tau;P) \geq c$, one can assess the size of shift required to push welfare below the value $c$. A similar exercise is performed in \cite{Spini2021} for aggregate policies, under the assumption that the distribution of $(Y_0,Y_1)|X$ is fixed and the distribution of $X$ varies over KL neighborhoods.
\end{remark}

\subsection{Implications}\label{sec:po.x.implications}

Before turning to empirical implementation, we first discuss some implications of the robust criteria derived in Proposition~\ref{prop:po.x.unbounded}. We highlight the different roles that treatment effect heterogeneity plays in the ranking of policies. We focus the discussion below on robust welfare, but the findings carry over to robust welfare gain also.

\subsubsection{Heterogeneity in CATEs}

Different patterns of heterogeneity in CATEs can have very different implications for the ranking of policies under distribution shifts. This is most easily seen when individual treatment effects $\Delta = Y_1 - Y_0$ conditional on characteristics are constant, i.e., $\Delta |(X = x) =_{a.s.} \delta(x)$ holds for all $x$ under $P$. We emphasize that our methods do not require this assumption; it is simply to facilitate the following discussion. Under constant treatment effects, the social welfare criterion (\ref{eq:social.welfare.criterion}) becomes 
\[
 \mr{W}(\tau;P) = \E_P[Y_0] + \E_P[\delta(X)|\tau(X) = 1] P(\tau(X) = 1)\,.
\]
Consider ranking a policy $\tau$ relative to a policy in which no one is treated. As $P(\tau(X) = 1) > 0$,  whether $\tau$ is preferred under criterion (\ref{eq:social.welfare.criterion}) is entirely determined by the sign of the average effect of treatment on the treated. In particular, the criterion is invariant to whether individuals with large negative treatment effects who are not treated under $\tau$ are ``almost'' treated in the sense that $h_1(X;\tau)$ is small. Similarly, it is invariant to whether treated individuals benefitting most from treatment are almost not treated. If so, such a policy may generalize poorly: there are distributions ``close'' to $P$ in which characteristics of a small mass of individuals with large negative or positive treatment effects are moved across the treatment/non-treatment frontier.

By contrast, the robust criterion from Proposition~\ref{prop:po.x.unbounded} becomes
\begin{multline*}
 \mr{RW}(\tau;P) = \E_P[Y_0] + \sup_{\eta \geq 1} \bigg( \E_P \left[ \min \left\{  \delta(X) + \eta h_1(X;\tau), 0 \right\}  | \tau(X) = 0 \right] P(\tau(X) = 0) \\
 + \E_P \left[ \min \left\{ \eta h_0(X;\tau) , \delta(X) \right\} | \tau(X) = 1  \right] P(\tau(X) = 1) - \eta \varepsilon  \bigg)\,.
\end{multline*}
Evidently, the robust criterion considers both the sign of treatment effects and their magnitudes relative to how close individuals are to the treatment/non-treatment frontier. ``Robust'' policies sort individuals by the size of their CATEs: individuals for whom $|\delta(X)|$ is small should be near the frontier, those for whom $|\delta(X)|$ is large should be far.

\begin{figure}[t]
\begin{subfigure}{.5\textwidth}
  \centering
  \includegraphics[width = 0.8\textwidth]{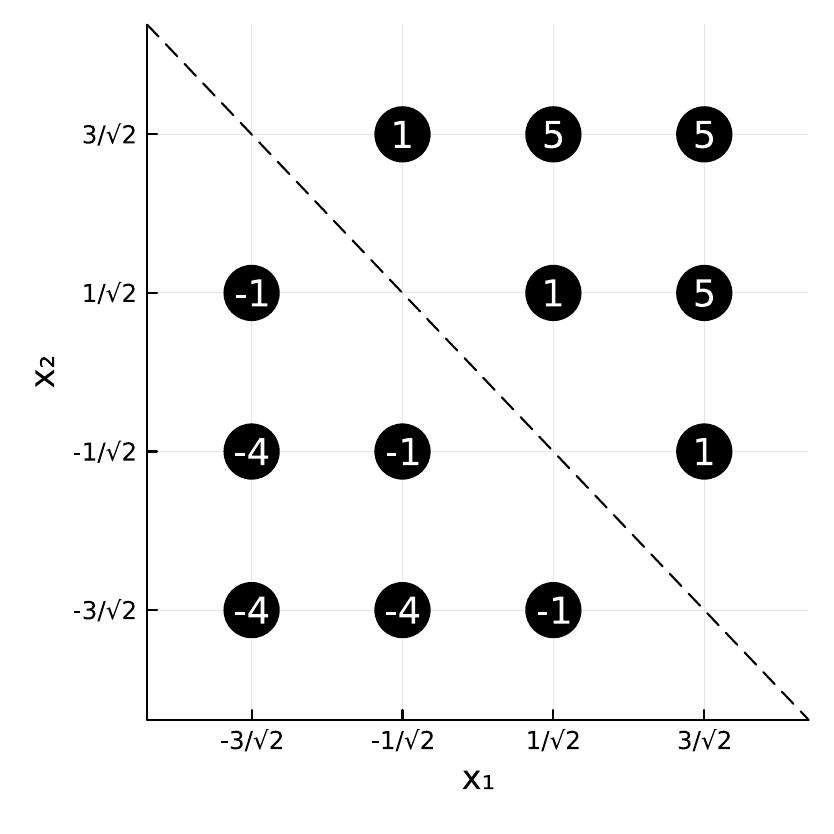}
  \caption{Population $P_a$} 
\end{subfigure}%
\begin{subfigure}{.5\textwidth}
  \centering
  \includegraphics[width = 0.8\textwidth]{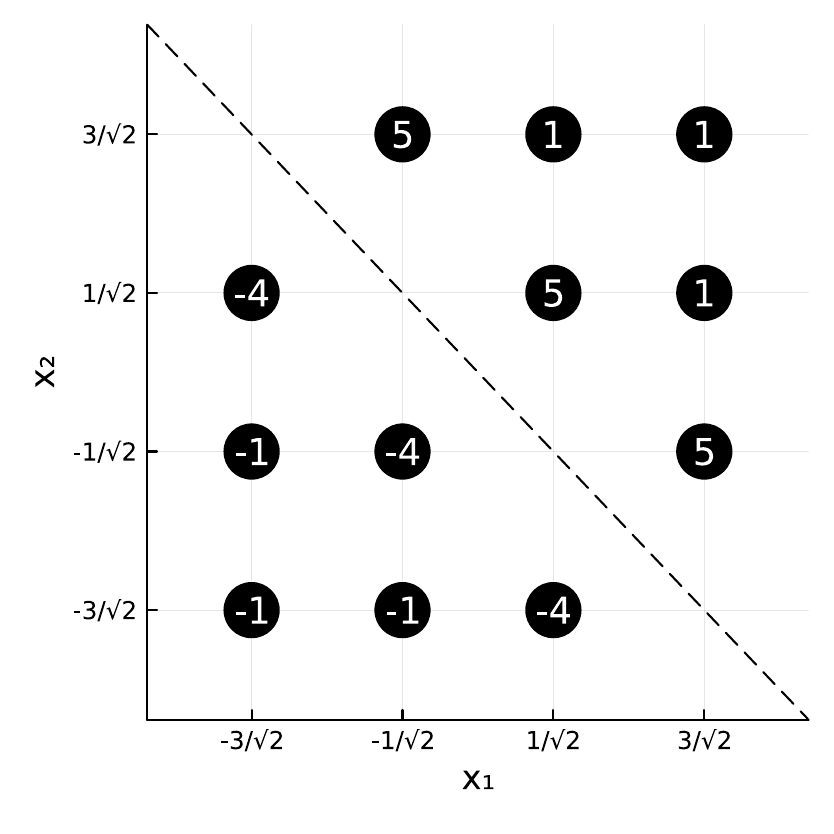}
  \caption{Population $P_b$}
\end{subfigure}
\caption{Two populations, different patterns of observable heterogeneity. \emph{Note:} each point corresponds to a mass of individuals with the same characteristics $(x_1,x_2)$. The number indicates the treated outcome $Y_1$, which is common to all individuals who share the same characteristics. All individuals in the population have $Y_0 = 0$. Individuals are uniformly distributed over the 12 mass points in each population.}
\label{fig:toy.example.P}
\end{figure}

We illustrate this with a numerical example. Consider the populations $P_a$ and $P_b$ plotted in Figure~\ref{fig:toy.example.P}. Individual treatment effects (conditional on $(x_1,x_2)$) are constant so there is no unobserved heterogeneity. Individuals with $x_1 + x_2 \geq 0$ benefit from treatment whereas those with $x_1 + x_2 < 0$ are adversely affected by it. The EWM policy in $P_a$ and $P_b$ is to treat individuals if and only if $x_1 + x_2 \geq 0$:\footnote{Here we refer to the welfare-optimal policy as the empirical welfare maximizing policy since the population distribution is known.}
\[
 \tau_{\text{\tiny EWM}}(x) = \mb I[ x_1 + x_2 \geq 0].
\]
The frontier between the treatment and non-treatment regions is represented by the dashed line in Figure~\ref{fig:toy.example.P}. This policy generates the same social welfare in $P_a$ and $P_b$,
\[
 \mr{W}(\tau_{\text{\tiny EWM}}; P_a) = \mr{W}(\tau_{\text{\tiny EWM}}; P_b),
\]
despite the different distributions of treatment effects in $P_a$ and $P_b$. A policy that treats everyone (``treat all'') generates welfare $0.25$ (the ATE) in both populations whereas a policy that treats no one (``treat none'') generates zero welfare since we have normalized $Y_0 = 0$ for all individuals in $P_a$ and $P_b$.

\begin{figure}[t]
\begin{subfigure}{.5\textwidth}
  \centering
  \includegraphics[width = \textwidth]{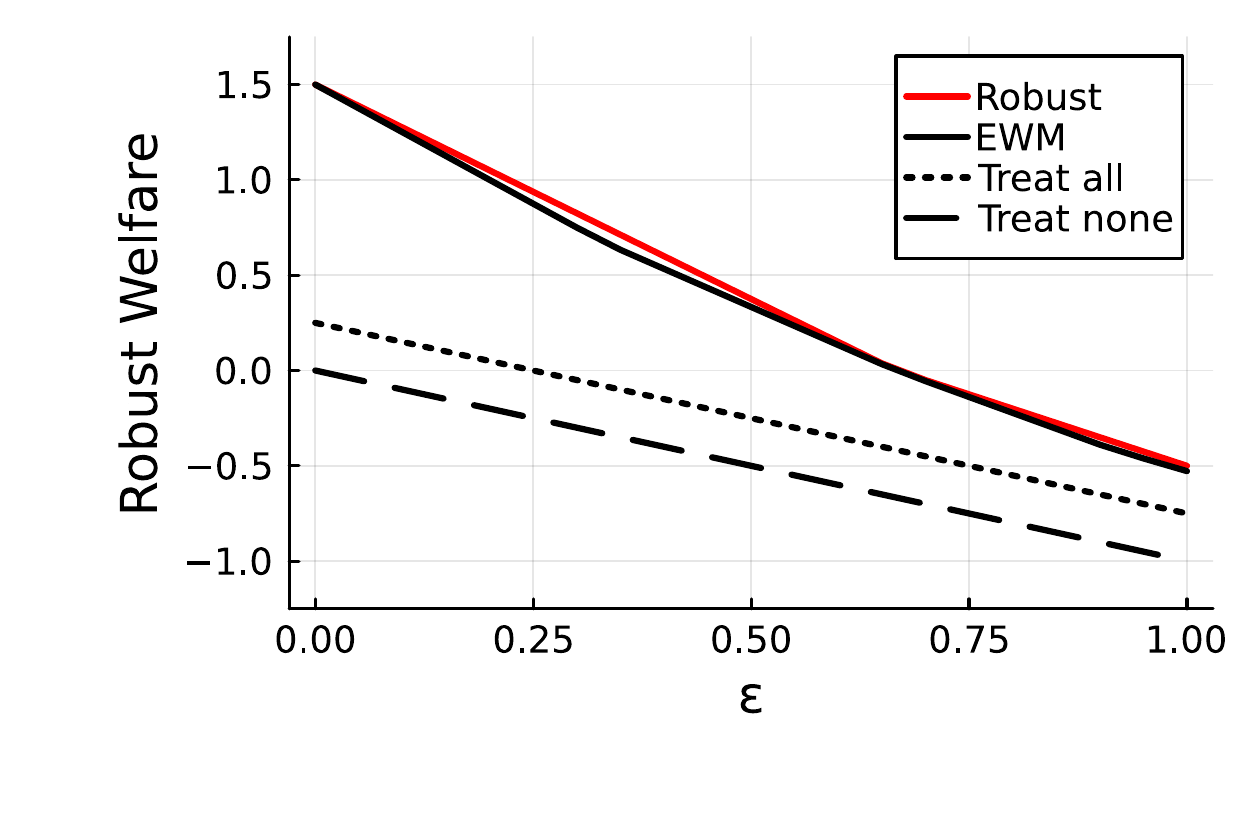}
  \caption{Population $P_a$} 
\end{subfigure}%
\begin{subfigure}{.5\textwidth}
  \centering
  \includegraphics[width = \textwidth]{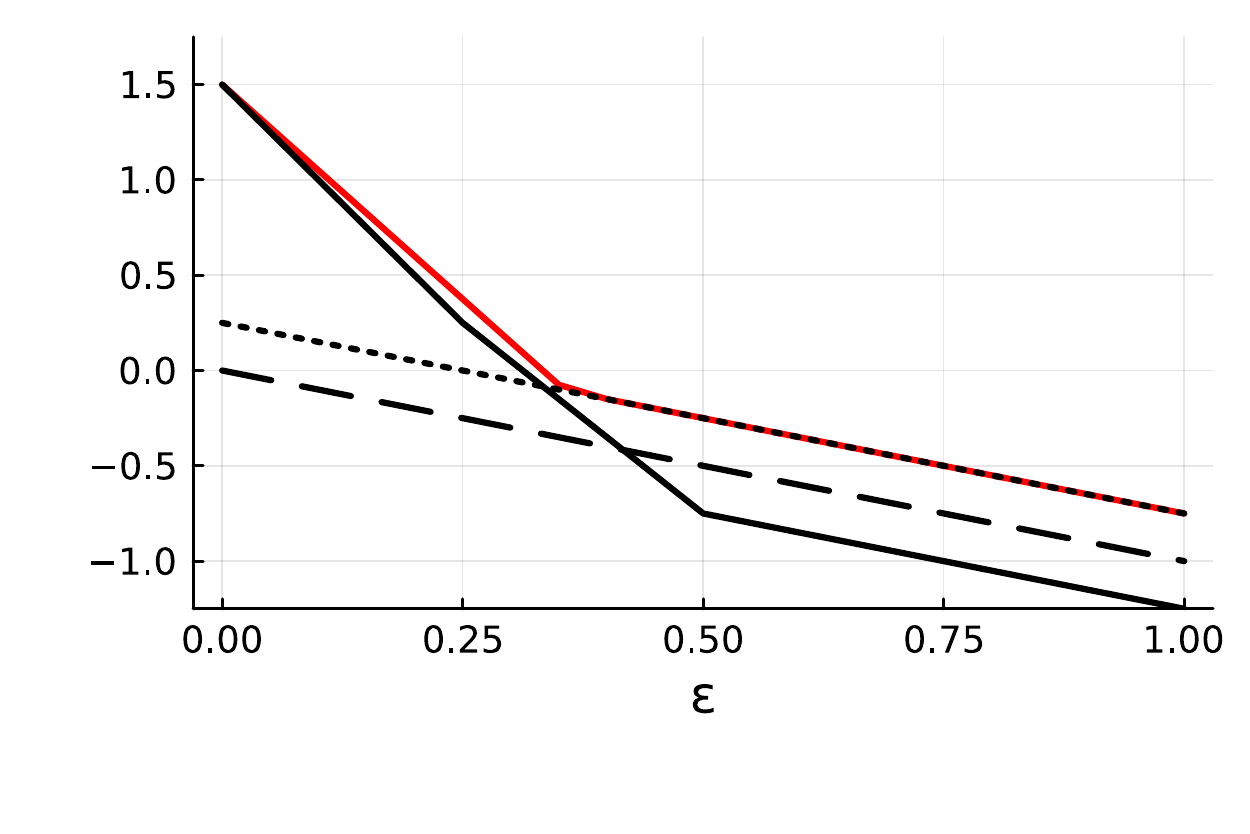}
  \caption{Population $P_b$}
\end{subfigure}
\caption{Robust welfare in $P_a$ and $P_b$ as a function of neighborhood size $\varepsilon$.}
\label{fig:toy.example.RW}
\end{figure}

Figure~\ref{fig:toy.example.RW} plots the robust welfare of the EWM, treat-all, and treat-none policies. 
The performance of the EWM policy is relatively insensitive to small perturbations of $P_a$. For instance with $\varepsilon = 0.5$, which represents twice the ATE in $P_a$, the robust welfare of the EWM policy is positive and exceeds that of the treat-all and treat-none policies. But the ranking of the three policies changes dramatically in $P_b$, where the EWM policy eventually performs \emph{worse} than both the treat-all and treat-none policies. For small $\varepsilon$, robust welfare of the EWM policy decreases at a rate of 5 times neighborhood size in $P_b$, double the rate of decrease in $P_a$. 

The EWM policy can be viewed as a special case of linear eligibility score policies in Example~\ref{ex:1}. We compute the robust policy that maximizes $\mr{RW}(\,\cdot\,;P)$ over this class for $P_a$ and $P_b$ and plot the resulting welfare in Figure~\ref{fig:toy.example.RW}. For $P_a$, the EWM policy is very robust: it performs almost as well as the robust policies which use different thresholds to determine treatment status. The robust policy is $\tau_{\text{\tiny Robust}}(x) = \mb I[0.314 + x_1 + x_2 \geq 0]$ for $\varepsilon < 2/3$, while for $2/3 < \varepsilon \leq 1$ the robust policy is $\tau_{\text{\tiny Robust}}(x) = \mb I[0.471 + x_1 + x_2 \geq 0]$. For $P_b$, the robust policy is $\tau_{\text{\tiny Robust}}(x) = \mb I[0.157 + x_1 + x_2 \geq 0]$ for $\varepsilon < 5/14$ while for $\varepsilon > 5/14$ the robust policy is no longer individualized but rather the treat-all policy. Here the EWM policy is reasonably robust to small shifts but performs much worse than the robust policy over shifts of size $\varepsilon \geq 0.5$. 

Overall, this example shows that individualized policies that leverage CATE heterogeneity within the experimental population may generalize poorly when there is a lot of heterogeneity near the treatment/non-treatment frontier.

\subsubsection{Unobserved Heterogeneity}\label{sec:unobs.het}

A second important consequence of Proposition~\ref{prop:po.x.unbounded} is that unobserved heterogeneity in treatment effects now plays a role. By contrast, the usual welfare criterion (\ref{eq:social.welfare.criterion}) is entirely invariant to unobserved heterogeneity. To see this, note that the robust criterion (\ref{eq:robust.po.x}) can be written
\begin{equation} \label{eq:robust.po.x.2}
  \E_P[Y_0] + \sup_{\eta \geq 1} \E_P \left[ \min \left\{  \Delta + \eta h_1(X;\tau) , \eta h_0(X;\tau)  \right\}   \right] - \eta \varepsilon \,.
\end{equation}
Consider populations $P_c$ and $P_d$ such that $(X,Y_0)$ has the same distribution in both, but $\Delta|(X = x) =_{a.s.} \delta(x)$ holds for all $x$ in $P_c$ whereas $\E_{P_d}[\Delta|X = x] = \delta(x)$ in $P_d$. Thus, the CATEs are the same in $P_c$ and $P_d$, but $P_d$ exhibits unobserved heterogeneity whereas $P_c$ does not. It follows from expression~(\ref{eq:robust.po.x.2}) that 
\[
 \mr{RW}(\tau;P_c) \geq \mr{RW}(\tau;P_d).
\]
The same is true for robust welfare gain. For intuition, nature's minimization over $Q$ transforms the objective from linear in $\Delta$ to concave in $\Delta$. 
Hence by Jensen's inequality, robust welfare is lower when there is unobserved heterogeneity. This is in contrast to the usual criteria (\ref{eq:social.welfare.criterion}) and (\ref{eq:welfare.gain.criterion}) which are invariant to unobserved heterogeneity:
\[
 \mr{W}(\tau;P_c) = \mr{W}(\tau;P_d)\,, \quad \quad \mr{WG}(\tau;P_c) = \mr{WG}(\tau;P_d)\,.
\]

\begin{figure}[t]
\begin{subfigure}{.5\textwidth}
  \centering
  \includegraphics[width = \textwidth]{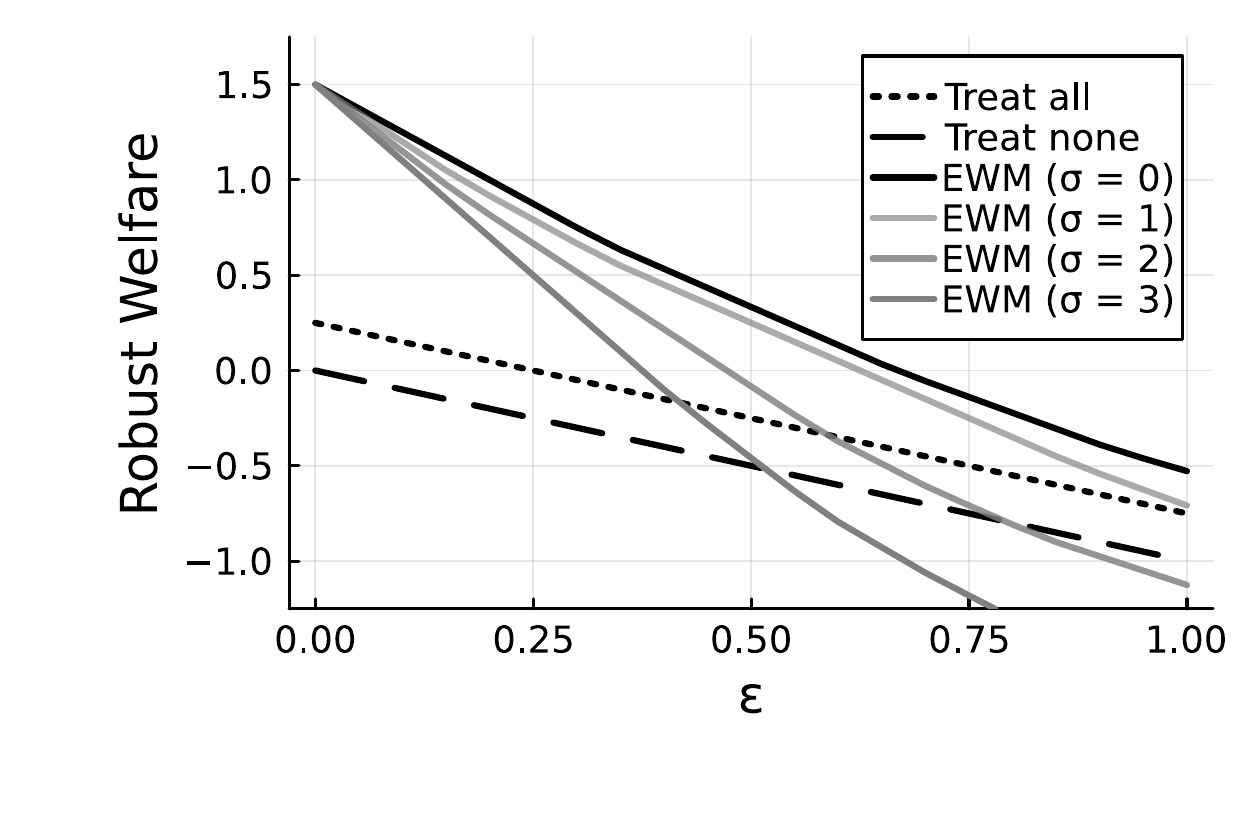}
  \caption{Performance of the EWM Policy} 
\end{subfigure}%
\begin{subfigure}{.5\textwidth}
  \centering
  \includegraphics[width = \textwidth]{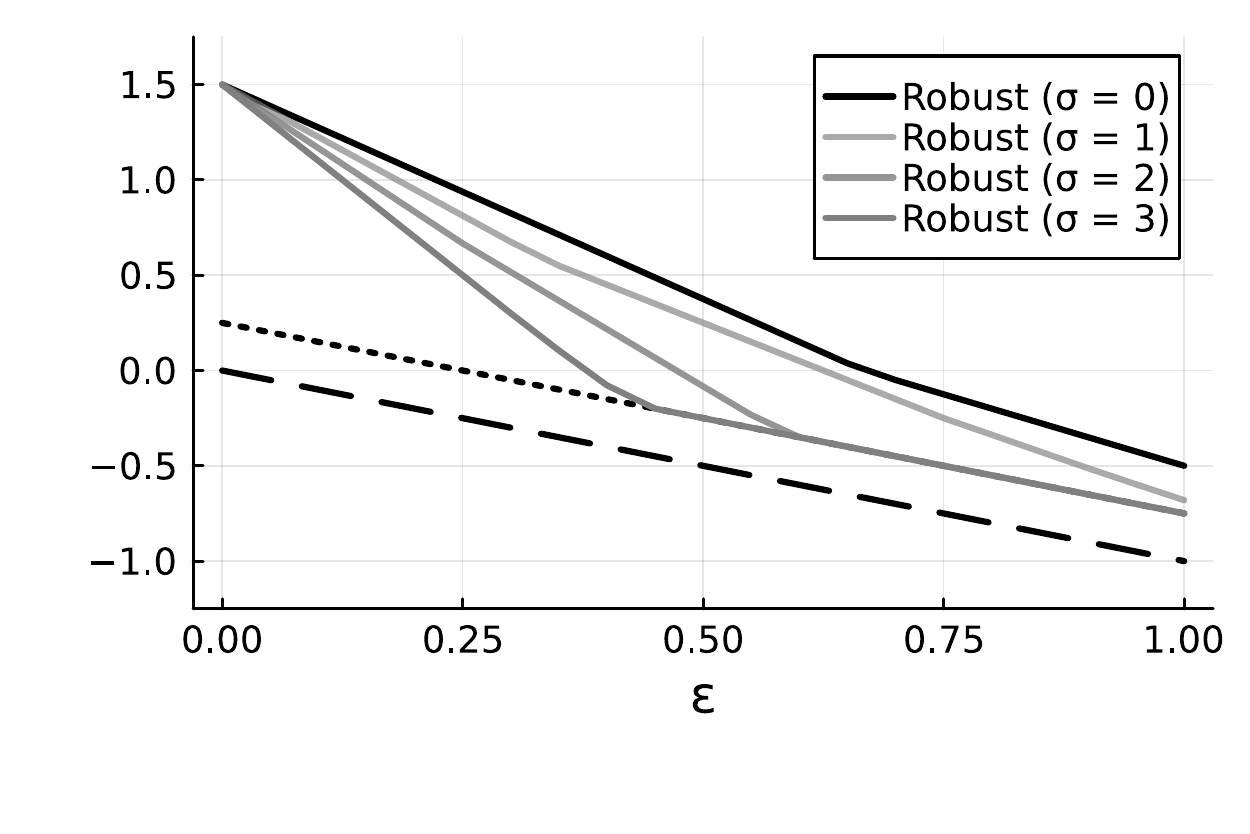}
  \caption{Performance of Robust Policies} 
\end{subfigure}
\caption{Robust welfare of policies for different values of the mean-preserving spread parameter $\sigma$. \emph{Note:} black lines correspond to $\sigma = 0$ (no unobserved heterogeneity), shaded lines to $\sigma \in \{1, 2, 3\}$ (darker for more unobserved heterogeneity).}
\label{fig:toy.example.mps}
\end{figure}

To see how unobserved heterogeneity plays a distinct role from heterogeneity in CATEs, consider the population $P_a$ described in Figure~\ref{fig:toy.example.P}(a). Individual treatment effects are degenerate (conditional on characteristics) in this population. We construct alternative populations with unobserved heterogeneity in treatment effects by adding independent mean-preserving spreads $\sigma W$ to individuals' treated outcomes, where $\sigma \geq 0$ and $P(W = 1) = P(W = -1) = \frac 12$. Hence, in these populations, the distributions of $X$ and CATEs are the same as in Figure~\ref{fig:toy.example.P}(a); all that changes is the variance of $Y_1$. Figure~\ref{fig:toy.example.mps}(a) plots the robust welfare of the EWM policy $\tau_{\text{\tiny EWM}}(x) = \mb I[ x_1 + x_2 \geq 0]$ for $\sigma \in \{0, 1, 2, 3\}$. When $\varepsilon = 0$, robust welfare is the same for all values of $\sigma$, reflecting the fact that the social welfare criterion (\ref{eq:social.welfare.criterion}) is invariant to unobserved heterogeneity. But as the degree of unobserved heterogeneity $\sigma$ and/or $\varepsilon$ increases, the EWM policy eventually performs worse than the treat-all and treat-none policies.

For each $\varepsilon$ and $\sigma$, we solve for the robust policy over the class of linear eligibility score policies. We plot the results in Figure~\ref{fig:toy.example.mps}(b). The solid black line corresponds to the robust policy for $\sigma = 0$ (this is the line plotted in red in Figure~\ref{fig:toy.example.RW}(a)), shaded lines correspond to robust policies with $\sigma > 0$. For small $\sigma$ and/or small $\varepsilon$, robust policies are all of the form $\tau_{\text{\tiny Robust}}(x) = \mb I[\beta_0 + x_1 + x_2 \geq 0]$, where $\beta_0 \geq 0$  depends on $\sigma$ and $\varepsilon$. For certain $\sigma$ and $\varepsilon$, the optimal $\beta_0 = 0$, in which case $\tau_{\text{\tiny Robust}} = \tau_{\text{\tiny EWM}}$. For other $\sigma$ and $\varepsilon$, the optimal $\beta_0 > 0$, in which case the robust policy has a larger treatment region than the EWM policy. For sufficiently large $\varepsilon$ and $\sigma$, the robust policy is no longer individualized but rather the treat-all policy.

\subsection{Empirical Strategies}\label{sec:po.x.empirical}

We now present strategies for estimating externally valid policies given a finite data set drawn from the experimental population. The challenge, seen from  (\ref{eq:robust.po.x.2}), is that the criterion involves the expectation of a nonlinear function of individual treatment effects $\Delta = Y_1 - Y_0$. Even in an experiment, the distribution of $(X,\Delta)$ is not identified without further assumptions. We first show how to estimate sharp bounds on robust welfare in the absence of identifying assumptions. We then show how to estimate robust welfare under different identifying assumptions. For brevity we confine most of the discussion that follows to the robust welfare criterion, though in view of (\ref{eq:equivalence}) the results carry over immediately to robust welfare gain.

\subsubsection{Sharp Bounds on Robust Welfare}\label{sec:bounds}

In this section we derive sharp bounds on the robust criteria and show how they may be estimated. The bounds depend only on the joint distributions of $(X,Y_0)$ and $(X,Y_1)$ in the experimental population $P$. These two distributions are identified under standard unconfoundedness (selection on observables) and overlap assumptions. They can also be identified from observational data under various assumptions.

Our approach follows \cite{HeckmanSmithClements1997}, \cite{Manski1997}, and in particular \cite{FanPark2010} and \cite{Stoye2010} (see also \cite{ImbensMenzel2021}). We seek to minimize and maximize $\mr{RW}(\tau;\tilde P)$ over all distributions $\tilde P$ for which the marginals for $(X,Y_0)$ and $(X,Y_1)$ are the same as under $P$. Different \emph{couplings} of the conditional distributions of $Y_0|X$ and $Y_1|X$ induce different distributions for $\Delta|X$. The expectation appearing in (\ref{eq:robust.po.x.2}) may be written as 
\[
  \E_P \left[ \E_P \left[ \left. \min \left\{  \Delta + \eta h_1(X;\tau) , \eta h_0(X;\tau)  \right\} \right| X \right]  \right].
\]
Because the inner expectation is taken over a concave function of $\Delta$, it suffices to consider second-order stochastic dominance (SSD) relations among distributions induced by different couplings. \cite{FanPark2010} show that perfect positive dependence (or rank invariance) of $Y_0|X$ and $Y_1|X$ produces a distribution of $\Delta|X$ that SSDs all distributions of $\Delta|X$ induced by other couplings. Similarly, perfect negative dependence (or rank reversal) produces a distribution of $\Delta|X$ that is SSD by all distributions induced by other couplings. The upper and lower bounds on robust welfare are therefore achieved under perfect positive and negative dependence, respectively: 
\[
 \mr{RW}(\tau; P^-) \leq \mr{RW}(\tau; P) \leq \mr{RW}(\tau; P^+) \quad \text{for all $\tau$},
\]
where $P^-$ and $P^+$ denote the joint distribution of $(X,Y_0,Y_1)$ under perfect negative and positive dependence of $Y_0|X$ and $Y_1|X$, respectively, for $P$-almost every $X$. An analogous ranking holds for robust welfare gain.

\begin{remark}\label{rmk:positive} \normalfont
The upper bound under perfect positive dependence is interesting in two respects. First, if rank invariance is a credible identifying assumption, then $P = P^+$ and $\mr{RW}(\tau; P) = \mr{RW}(\tau; P^+)$ and $\mr{RWG}(\tau; P) = \mr{RWG}(\tau; P^+)$ for all policies $\tau$. In this case, one can estimate policies that are robust to shifts in outcomes and characteristics by maximizing the empirical robust welfare criterion (\ref{eq:robust.po.x.empirical}) introduced below or its modification for robust welfare gain.

Second, the upper bound can serve as a stress test of a given policy, say $\hat \tau$: if the robust welfare (gain) of $\hat \tau$ under $P^+$ is bad, then robust welfare (gain) under $P$ must be even worse. The empirical robust welfare criterion (\ref{eq:robust.po.x.empirical}) provides an estimate of $\mr{RW}(\hat \tau; P^+)$, which may be used to evaluate the robustness of estimated policies.
\end{remark}

\begin{remark} \normalfont
The lower bound is also interesting as it gives a welfare guarantee under both adversarial shifts between the experimental and target populations and adversarial couplings in the experimental population. The empirical robust welfare  (gain) criterion under perfect negative dependence introduced below can be used to estimate robust policies which guard against both adversaries.
\end{remark}

\bigskip

\paragraph{Empirical Implementation.}
Suppose we observe a random sample $(X_i,Y_i,D_i)_{i=1}^n$, where $D_i$ is a binary treatment indicator, $Y_i = D_i Y_{1i} + (1-D_i)Y_{0i}$, and  the unconfoundedness condition 
\begin{equation} \label{eq:unconfoundedness}
 (Y_0,Y_1) \perp \!\!\! \perp D \,|\, X
\end{equation}
and overlap condition 
\begin{equation} \label{eq:overlap}
 0 < \Pr[D=1|X = x] < 1
\end{equation}
for all $x$ in the support of $X$, both hold. Then both $F_0(y|x) = P(Y_0 \leq y|X = x)$ and $F_1(y|x) = P(Y_1 \leq y|X = x)$ are identified:
\begin{equation} \label{eq:distribution.exogenous}
 F_d(y|x) = \Pr(Y \leq y | X = x, D = d), \quad d = 0,1.
\end{equation}
Standard nonparametric methods can be used to estimate $F_0$ and $F_1$ based on (\ref{eq:distribution.exogenous}). If unconfoundedness fails, then $F_0(y|x)$ and $F_1(y|x)$ may be identified using instrumental variables---see, e.g., \cite{VuongXu2017} and references therein.

To simplify exposition, suppose that $Y_0$ and $Y_1$ are continuously distributed conditional on $X$. Given $F_0$ and $F_1$, counterfactual mappings may be constructed as
\[
 \phi_x^+(\cdot) = F_1^{-1}(F_0(\,\cdot\,|x)|x) \,, \quad \quad \phi_x^-(\cdot) = F_1^{-1}(1-F_0(\,\cdot\,|x)|x) \,.
\]
The function $\phi_x^+$ maps an individual's untreated outcome to their treated outcome under perfect positive dependence; its inverse maps treated outcomes to untreated outcomes. The function $\phi_x^-$ and its inverse give the same counterfactual mappings under perfect negative dependence. Individual $i$'s treatment effects under perfect positive and negative dependence are therefore
\begin{equation*}
 \begin{aligned}
 \Delta_i^{\!+} & = D_i \left( Y_i - (\phi_{X_i}^+)^{-1}(Y_i) \right) + (1 - D_i) \left( \phi_{X_i}^+(Y_i) - Y_i \right) , \\
 \Delta_i^{\!-} & = D_i \left( Y_i - (\phi_{X_i}^-)^{-1}(Y_i) \right) + (1 - D_i) \left( \phi_{X_i}^-(Y_i) - Y_i \right) .
 \end{aligned}
\end{equation*}
Given estimates $\hat F_0$ and $\hat F_1$, we may construct estimates $\hat \phi_x^+$ and $\hat \phi_x^-$ of $\phi_x^+$ and $\phi_x^-$. We may then estimate individual treatment effects under both dependence assumptions using
\begin{equation*}
 \begin{aligned}
 \hat \Delta_i^{\!+} & = D_i \left( Y_i - (\hat \phi_{X_i}^+)^{-1}(Y_i) \right) + (1 - D_i) \left(\hat \phi_{X_i}^+(Y_i) - Y_i \right) , \\
 \hat \Delta_i^{\!-} & = D_i \left( Y_i - (\hat \phi_{X_i}^-)^{-1}(Y_i) \right) + (1 - D_i) \left(\hat \phi_{X_i}^-(Y_i) - Y_i \right) .
 \end{aligned}
\end{equation*}
Finally, given an estimate $\bar Y_0$ of $\E_P[Y_0]$, we construct an \emph{empirical robust welfare} criterion under perfect positive dependence as follows:
\begin{equation}\label{eq:robust.po.x.empirical}
 \mr{ERW}_n(\tau;P^+)  = \bar Y_0 +  \sup_{\eta \geq 1} \frac{1}{n} \sum_{i=1}^n \min\left\{  \hat \Delta_i^{\!+} + \eta h_1(X_i; \tau), \eta h_0(X_i; \tau) \right\}  - \eta\varepsilon  .
\end{equation}
The criterion $\mr{ERW}_n(\tau;P^-)$ for perfect negative dependence is constructed analogously, replacing $\hat \Delta_i^+$ with $\hat \Delta_i^-$ in the above expression.  In view of (\ref{eq:equivalence}), an empirical counterpart to robust welfare gain for $\tau \not \equiv 0$ can be constructed by setting $\bar Y_0 = 0$ in (\ref{eq:robust.po.x.empirical}), and setting empirical robust welfare gain to zero for $\tau \equiv 0$. As such, Propositions~\ref{prop:robust.po.x.empirical.bounds} and~\ref{prop:robust.po.x.empirical.bounds.rate} below apply equally to robust welfare gain.

In practice, the optimization over $\eta$ may be efficiently performed using linear programming. For scalars $a_i,b_i,c_i$, $i = 1,\ldots,n$,  we have
\begin{equation}\label{eq:robust.po.x.empirical.1.lp}
\begin{aligned}
 \sup_{\eta \geq 1} \frac{1}{n} \sum_{i=1}^n \min\left\{  a_i + \eta b_i , \eta c_i \right\} - \eta \varepsilon 
 & = \sup_{\eta, (t_i)_{i=1}^n} \frac{1}{n} \sum_{i=1}^n t_i - \eta \varepsilon , \quad  \mbox{subject to} \quad \eta \geq 1 ,  \\
 & \quad \quad t_i \leq a_i + \eta b_i  , \quad t_i \leq \eta c_i , \quad i = 1,\ldots,n.
\end{aligned}
\end{equation}

\begin{proposition}\label{prop:robust.po.x.empirical.bounds}
Suppose that the conditions of Proposition~\ref{prop:po.x.unbounded} hold, $\mc X$ is a bounded subset of $\mb R^d$, $\frac{1}{n} \sum_{i=1}^n  |\hat \Delta^{\!+}_i - \Delta^{\!+}_i| \to_p 0$, and $\bar Y_0 \to_p \E_P[Y_0]$. Then
\begin{enumerate}[nosep, left=0pt]
\item $\sup_{\tau \in \mc T} |\mr{ERW}_n(\tau;P^+) - \mr{RW}(\tau;P^+)| \to_p 0$; 
\item $\sup_{\tau \in \mc T}\mr{RW}(\tau;P^+) - \mr{RW}(\hat \tau;P^+) \to_p 0$ for any maximizer $\hat \tau$ of $\mr{ERW}_n(\,\cdot\,;P^+)$.
\end{enumerate}
Moreover, the convergence in parts~1. and~2.~holds uniformly for $\varepsilon \geq \ul \varepsilon$ for any arbitrarily small $\ul \varepsilon > 0$.
An analogous result holds for $\mr{ERW}_n(\tau;P^-)$.
\end{proposition}

\begin{remark}\normalfont
Proposition~\ref{prop:robust.po.x.empirical.bounds} holds for any (and hence all) classes of policies $\mc T$. The cost of this generality is the requirement of a bounded characteristic space $\mc X$, which is used to restrict the complexity of  $\{\min\{ \Delta^{\!\pm} + \eta h_1(X;\tau), \eta h_0(X;\tau)\} : \tau \in \mc T\}$. Only characteristics that are arguments of $\tau$ need to be bounded. If there are additional characteristics introduced, e.g., to ensure unconfoundedness is credibly satisfied but these characteristics do not appear as arguments of $\tau$, then they may be unbounded.
\end{remark}

\begin{remark}\normalfont
Solving for the robust policy $\hat \tau$ under positive or negative dependence requires maximizing $\mathrm{ERW}_n(\tau; P^+)$ or $\mathrm{ERW}_n(\tau;P^-)$ over $\tau \in \mathcal T$. As (\ref{eq:robust.po.x.empirical.1.lp})  shows, the objective function is the optimal value of a linear program. Gradient-based optimizers may be unreliable because the optimal value of linear programs can depend non-smoothly on parameters (see, e.g., \cite{MilgromSegal2002}). Moreover, there may be multiple local optima. We therefore recommend computing $\hat \tau$ via a grid search, which is feasible as $\mathcal T$ is typically a simple parametric family (cf.~Examples~\ref{ex:1}--\ref{ex:3}). In the numerical examples in Section~\ref{sec:po.x.implications} and the application in Section~\ref{sec:empirical}, we first ran a grid search to find a set of parameter values that approximately maximized the robust welfare criterion, then used these as starting values in a gradient-free optimization.

 In the related problem of classification/empirical welfare maximization, computation can be simplified by replacing the original objective function by a ``surrogate'', i.e., a smooth convex function for which the optimizer is the same (see, e.g., \cite{BJM2006}). It would be interesting to explore whether surrogates can be used to simplify computation for robust policy learning problems.
\end{remark}

It is possible to relax boundedness of $\mc X$ and derive convergence rates under conditions on $\mc T$ and the estimators $\hat \Delta^{\!\pm}$ and $\bar Y_0$.
Here we give one such result. Each $\tau \in \mc T$ is identified with a decision set $C$ such that $\tau(x) = \mb I [x \in C]$. Let $\mc C$ be the set of all decision sets for $\tau \in \mc T$. A standard assumption is that $\mc C$ is a \emph{VC class} (i.e., has finite VC dimension),\footnote{We refer the reader to Chapter 2.6 of \cite{vanderVaartWellner} for terminology.} which means $\mc T$ has the same complexity as a parametric class. We impose a related condition. 
Let $\mc C^c = \{C^c : C \in \mc C\}$ and let $\mc C^* = \{ \{(x,t) \in \mathcal X \times \mathbb R_+ : \inf_{\tilde x \in C} \|x - \tilde x\| \leq t / \eta\}:  C \in \mc C \cup \mc C^c, \eta \geq 1\}$. We require that $\mc C^*$ is a VC class, though we allow its VC dimension $v_n$ to grow with the sample size. In what follows, the condition $v_n \gtrsim 1$ means that $v_n$ is bounded below by a positive constant for all $n$ sufficiently large.

\begin{proposition}\label{prop:robust.po.x.empirical.bounds.rate}
Suppose that the conditions of Proposition~\ref{prop:po.x.unbounded} hold,  $\E_P[Y_d^2] < \infty$ for $d = 0,1$, $\mc C^*$ is a VC class of dimension $v_n \gtrsim 1$, and there are positive constants $a_n, b_n$ such that $\frac{1}{n} \sum_{i=1}^n |\hat \Delta_i^{\!+} - \Delta_i^{\!+}| = O_p(a_n)$ and $|\bar Y_0 - \E_P[Y_0]| = O_p(b_n)$. Then with $c_n = \max\{a_n, b_n, (v_n / n)^{1/2}\}$,
\begin{enumerate}[nosep, left=0pt]
\item $\sup_{\tau \in \mc T} |\mr{ERW}_n(\tau;P^+) - \mr{RW}(\tau;P^+)| = O_p( c_n)$; 
\item $\sup_{\tau \in \mc T}\mr{RW}(\tau;P^+) - \mr{RW}(\hat \tau;P^+) = O_p( c_n)$ for any maximizer $\hat \tau$ of $\mr{ERW}_n(\,\cdot\,;P^+)$.
\end{enumerate}
Moreover, the convergence rates in parts~1. and~2.~hold uniformly for $\varepsilon > 0$.
An analogous result holds for $\mr{ERW}_n(\tau;P^-)$.
\end{proposition}

\begin{remark} \normalfont
The convergence rate $b_n$ will typically be $O(n^{-1/2})$, leading to an overall rate of $c_n = O(\max\{a_n, (v_n/n)^{1/2}\})$. We conjecture that it may be possible to reduce or remove the $a_n$ term to recover an $O( (v_n/n)^{1/2})$ rate (i.e., the minimax rate for welfare maximization over a VC class of policies of dimension $v_n$) using a doubly/locally robust construction similar to that used by \cite{AtheyWager2021} for empirical welfare maximization. The situation here is more complicated, however, as $\Delta$ enters the criterion non-smoothly. 
\end{remark}

\subsubsection{Identifying Assumption: No Unobserved Heterogeneity}\label{sec:constant.treatment}

A simple approach to identifying $P$ is to assume treatment effects are constant conditional on $X$, i.e., 
\begin{equation}\label{eq:po.degenerate}
 \Delta |(X = x) =_{a.s.} \delta(x)\,
\end{equation}
for some function $\delta : \mc X \to \mb R$. Under this assumption, treatment effects can vary across individuals with different characteristics but are homogeneous for individuals with the same characteristics. While this assumption implies perfect positive dependence, it simplifies empirical implementation as the problem of estimating the counterfactual mapping $\phi_x^+$ is replaced by the (simpler) problem of estimating $\delta$.

\bigskip

\paragraph{Empirical Implementation.}
Suppose we observe a random sample $(X_i,Y_i,D_i)_{i=1}^n$ where $D_i$ is a binary treatment indicator and $Y_i = D_i Y_{1i} + (1-D_i)Y_{0i}$. Then $\delta$ is identified under the unconfoundedness condition (\ref{eq:unconfoundedness}) and the overlap condition (\ref{eq:overlap}). We may estimate $\delta$ using a variety of nonparametric regression techniques. 
Alternatively, suppose we observe $(X_i,Y_i,D_i,Z_i)_{i=1}^n$ where $Z_i$ is an instrumental variable satisfying appropriate regularity conditions (see, e.g., \cite{Abadie2003}). By (\ref{eq:po.degenerate}), $\delta$ is identified as the ratio
\[
 \delta(x) = \frac{\mr{Cov}(Y_i,Z_i|X_i = x)}{\mr{Cov}(D_i,Z_i|X_i = x)}
\]
and may be estimated using  nonparametric instrumental variables methods.

In either case, given estimators $\hat \delta$ of $\delta$ and $\bar Y_0$ of $\E_P[Y_0]$, we take $\hat \Delta_i \equiv \hat \delta(X_i)$ to be our estimate of $\Delta_i \equiv \delta(X_i)$ and form the empirical robust welfare criterion 
\begin{equation}\label{eq:constant}
 \mr{ERW}_n(\tau;P) = 
 \bar Y_0 +  \sup_{\eta \geq 1} \frac{1}{n} \sum_{i=1}^n \min\left\{  \hat \Delta_i + \eta h_1(X_i;\tau) \,, \eta h_0(X_i;\tau) \right\}  - \eta\varepsilon .
\end{equation}
As before, the optimization over $\eta$ can be efficiently implemented via linear programming (see equation~(\ref{eq:robust.po.x.empirical.1.lp})). We can estimate policies that are robust to shifts in outcomes and characteristics by maximizing this criterion with respect to $\tau \in \mc T$. Asymptotic properties of the estimated policy $\hat \tau$ may be established under a variety of regularity conditions. The following high-level results allow for both experimental and observational data.

\begin{proposition}\label{prop:robust.po.x.empirical.1}
Suppose that the conditions of Proposition~\ref{prop:po.x.unbounded} hold, condition (\ref{eq:po.degenerate}) holds, $\mc X$ is a bounded subset of $\mb R^d$, $\frac{1}{n} \sum_{i=1}^n  |\hat \Delta_i - \Delta_i| \to_p 0$, and $\bar Y_0 \to_p \E_P[Y_0]$. Then
\begin{enumerate}[nosep, left=0pt]
\item $\sup_{\tau \in \mc T} |\mr{ERW}_n(\tau;P) - \mr{RW}(\tau;P)| \to_p 0$; 
\item $\sup_{\tau \in \mc T}\mr{RW}(\tau;P) - \mr{RW}(\hat \tau;P) \to_p 0$ for any maximizer $\hat \tau$ of $\mr{ERW}_n(\,\cdot\,;P)$.
\end{enumerate}
Moreover, the convergence in parts~1. and~2.~holds uniformly for $\varepsilon \geq \ul \varepsilon$ for any arbitrarily small $\ul \varepsilon > 0$.
\end{proposition}

For the next result, we again identify each $\tau \in \mc T$ with a decision set $C$ such that $\tau(x) = \mb I[x \in C]$ and impose a complexity condition from Section~\ref{sec:bounds} on the  sets.

\begin{proposition}\label{prop:robust.po.x.empirical.1.rate}
Suppose that the conditions of Proposition~\ref{prop:po.x.unbounded} hold, condition (\ref{eq:po.degenerate}) holds,  $\E_P[Y_d^2] < \infty$ for $d = 0,1$, $\mc C^*$ is a VC class of dimension $v_n \gtrsim 1$, and there are positive constants $a_n, b_n$ such that $\frac{1}{n} \sum_{i=1}^n |\hat \Delta_i - \Delta_i| = O_p(a_n)$ and $|\bar Y_0 - \E_P[Y_0]| = O_p(b_n)$. Then with $c_n = \max\{a_n, b_n, (v_n / n)^{1/2}\}$,
\begin{enumerate}[nosep, left=0pt]
\item $\sup_{\tau \in \mc T} |\mr{ERW}_n(\tau;P) - \mr{RW}(\tau;P)| = O_p( c_n)$; 
\item $\sup_{\tau \in \mc T}\mr{RW}(\tau;P) - \mr{RW}(\hat \tau;P) = O_p( c_n)$ for any maximizer $\hat \tau$ of $\mr{ERW}_n(\,\cdot\,;P)$.
\end{enumerate}
Moreover, the convergence rates in parts~1. and~2.~hold uniformly for $\varepsilon > 0$.
\end{proposition}

\subsubsection{Identifying Assumption: Conditional Independence}

An alternative identifying assumption is to posit the existence of a random variable $V$ such that all dependence between $Y_0$ and $Y_1$ given $X$ comes through $V$:
\begin{equation}\label{eq:conditional.independence}
 Y_0 \perp \!\!\! \perp Y_1 \,|\, X, V
\end{equation}
\citep[Section 2.5.1]{AbbringHeckman2007}, and that a suitably modified version of unconfoundedness holds:
\begin{equation}\label{eq:conditional.unconfoundedness}
 (Y_0,Y_1) \perp \!\!\! \perp D \,|\, (X, V)\,.
\end{equation} 
The variables $V$ may coincide with those in $X$, or may be distinct. For example, $Y_0$ and $Y_1$ may be conditionally independent given $X$ and a latent common factor $F$, and $V$ could be a perfect proxy for $F$ constructed from $X$ and other observables. A special case is group fixed effects, where potential outcomes (and possibly the assignment mechanism) have a group-specific component $\alpha_V$. Using a linear model for simplicity, we might have
\[
\begin{aligned}
 Y_0 & = X'\beta_0 + \lambda_0 \alpha_{V} + u_0, \\
 Y_1 & = X'\beta_1 + \lambda_1 \alpha_{V} + u_1, 
\end{aligned}
\]
with $V$ denoting the individual's group membership (assumed known) and where $u_0$ and $u_1$ are conditionally independent given $(X,V)$.

Suppose the analyst observes a random sample $(X_i,Y_i,D_i,V_i)_{i=1}^n$ in which (\ref{eq:conditional.independence}) and (\ref{eq:conditional.unconfoundedness}) hold and a suitably modified version of the overlap condition (\ref{eq:overlap}) holds. The conditional CDFs $F_0(\cdot|x,v)$ and $F_1(\cdot|x,v)$ are then nonparametrically identified as the conditional CDFs of $Y$ given $D = d,X = x,V=v$ for $d = 0,1$, respectively. Assumption (\ref{eq:conditional.independence}) then permits identification of the robust welfare criterion under $P$. Note the conditional CDF $G_{\eta,\tau}(\cdot|x,v)$ of $\min\left\{ Y_0 + \eta h_0(X;\tau)\,, Y_1 + \eta h_1(X;\tau) \right\} $ given $X = x$ and $V = v$ is
\[
 G_{\eta,\tau}(z|x,v) = 1 - (1 - F_0(z-\eta h_0(x;\tau)|x,v))(1 - F_1(z-\eta h_1(x;\tau)|x,v)),
\]
where $F_0$ and $F_1$ are the conditional CDFs of $Y_0$ and $Y_1$ given $X = x$ and $V = v$. It follows by iterated expectations that 
\[
 \E_P \left[  \min\left\{ Y_0 + \eta h_0(X;\tau)\,, Y_1 + \eta h_1(X;\tau) \right\} \right] 
 = \int \int z \, \mr d G_{\eta,\tau}(z|x,v) \, \mr d P_{X,V}(x,v)\,,
\]
where $P_{X,V}$ is the distribution of $(X,V)$ in the experimental population.

 Given estimators $\hat F_0(y|x,v)$ and $\hat F_1(y|x,v)$, we estimate $G_{\eta,\tau}$ using
\[
 \hat G_{\eta,\tau}(z|x,v) = 1 - (1 - \hat F_0(z-\eta h_0(x;\tau)|x,v))(1 - \hat F_1(z-\eta h_1(x;\tau)|x,v))\,.
\]
A useful feature of $\hat G_{\eta,\tau}$ is that it is uniformly consistent in $(z,\eta,\tau)$ whenever $\hat F_0(y|x,v)$ and $\hat F_1(y|x,v)$ are uniformly consistent in $y$. 
We choose $\hat \tau$ by maximizing the empirical robust welfare criterion
\begin{equation}\label{eq:kernel}
 \mr{ERW}_n(\tau;P) = \sup_{\eta \geq 1} \frac{1}{n} \sum_{i=1}^n \int z \, \mr d \hat G_{\eta,\tau}(z|X_i,V_i)  - \eta\varepsilon 
\end{equation}
with respect to $\tau \in \mc T$. The following result establishes consistency. Let $\bar F_d(y) = \frac{1}{n} \sum_{i=1}^n \hat F_{d}(y|X_i,V_i)$ for $d = 0,1$, let $a,A$ be finite positive constants, and let wpa1 denote ``with probability approaching one''.

\begin{proposition}\label{prop:robust.po.x.empirical.3}
Suppose that the conditions of Proposition~\ref{prop:po.x.unbounded} hold, the conditional independence condition (\ref{eq:conditional.independence}) holds, $\mc X$ is bounded, and $\int |y|^{1+a} \mr d \bar F_d(y) \leq A$ wpa1 and $\frac{1}{n} \sum_{i=1}^n \sup_y |\hat F_d(y|X_i,V_i) - F_d(y|X_i,V_i)| \to_p 0$ for $d = 0,1$. Then
\begin{enumerate}[nosep, left=0pt]
\item $\sup_{\tau \in \mc T} |\mr{ERW}_n(\tau;P) - \mr{RW}(\tau;P)| \to_p 0$; 
\item $\sup_{\tau \in \mc T}\mr{RW}(\tau;P) - \mr{RW}(\hat \tau;P) \to_p 0$ for any maximizer $\hat \tau$ of $\mr{ERW}_n(\,\cdot\,;P)$.
\end{enumerate}
Moreover, the convergence in parts~1. and~2.~holds uniformly for $\varepsilon \geq \ul \varepsilon$ for any arbitrarily small $\ul \varepsilon > 0$.
\end{proposition}

\subsection{Extensions and Complements}\label{sec:extensions}

\subsubsection{Shutting Down Unobserved Heterogeneity}\label{sec:nohet}

As discussed in Section~\ref{sec:po.x.implications}, heterogeneity in CATEs and unobserved heterogeneity both play distinct yet important roles in shaping generalizability. To better understand the separate roles they play, we can shut down the latter channel as follows.\footnote{We are grateful to an anonymous referee for this suggestion.} We begin by writing the usual welfare objective (\ref{eq:social.welfare.criterion}) as
\[
 \mr{W}(\tau;P) = \E_P[ S \, \tau(X) + M] ,
\]
where $(S,M)|(X = x) =_{a.s.} (\delta(x), \mu_0(x))$ under $P$, with $\delta(x) = \E[ \Delta|X = x]$ and $\mu_0(x) = \E[Y_0|X = x]$. Suppose we wish to consider robustness with respect to shifts in the distribution of $(S, M, X)$. We constrain the class of distributions to Wasserstein neighborhoods centered at $P$ and defined using the metric
\[
 d((\delta,m,x),(\tilde \delta, \tilde m, \tilde x)) = |\delta - \tilde \delta| + |m - \tilde m| + \|x - \tilde x\|.
\]
By similar arguments to Proposition~\ref{prop:po.x.unbounded}, we arrive at the robust welfare objective
\begin{equation}\label{eq:criterion.nohet}
 \mr{RW}(\tau;P) = \E_P[Y_0] + \sup_{\eta \geq 1} \E_P \left[ \min \left\{ \eta h_0(X;\tau) , \delta(X) + \eta h_1(X;\tau) \right\}   \right] - \eta \varepsilon .
\end{equation}
This is identical to the robust welfare formulation from Section~\ref{sec:constant.treatment}, except we now interpret $\delta(x)$ as the CATE. Given estimates $\hat \delta$ of $\delta$ and $\bar Y_0$ of $\E_P[Y_0]$, we can estimate policies that are robust to this class of distribution shifts by maximizing criterion~(\ref{eq:constant}) with $\hat \Delta_i = \hat \delta(X_i)$. Theoretical properties of the estimated policies follow by similar arguments to Propositions~\ref{prop:robust.po.x.empirical.1} and~\ref{prop:robust.po.x.empirical.1.rate}, without requiring~\eqref{eq:po.degenerate}.

In related work, \cite{Kido2022} uses a sequential neighborhood construction in which the distribution of $(Y_0,Y_1)|X$ can vary over a Wasserstein neighborhood, then the marginal distribution of $X$ can shift over an $f$-divergence neighborhood. That construction likewise eliminates the influence of unobserved heterogeneity. Moreover, it results in an objective function that is essentially equivalent to that of \cite{MoQiLiu}, who fix the conditional distribution of $(Y_0,Y_1)|X$ between the experimental and target populations, but allow the marginal distribution of $X$ to vary over $f$-divergence neighborhoods. These approaches similarly produce a criterion function involving an expectation over a concave function of the CATE that is optimized with respect to a Lagrange multiplier representing the neighborhood constraint.

\bigskip

\paragraph{Example: Joint Shifts or CATE-based Shifts?}

The construction here in essence replaces individual treatment effects with CATEs, whereas our earlier approach allows the joint distribution of $(X,Y_0,Y_1)$ to shift freely. 

To understand the differences between these approaches, consider the numerical example presented in Section~\ref{sec:unobs.het}. There we start with a given DGP with no unobserved heterogeneity, then add mean-preserving spreads $\sigma W$ to the treated outcome $Y_1$, where $\sigma \geq 0$ is a scale parameter and $P(W = 1) = P(W = -1) = \frac 12$. Increasing $\sigma$ leaves the CATE unchanged but increases unobserved heterogeneity. 

Under criterion (\ref{eq:criterion.nohet}), the conditional distribution of treatment effects around the CATE (i.e., unobserved heterogeneity) is irrelevant. Thus, for any $\sigma \geq 0$, criterion (\ref{eq:criterion.nohet}) yields the same value of robust welfare and therefore the same ranking over treatment policies. By contrast, Figure~\ref{fig:toy.example.mps} shows that this is not so for the earlier criterion (\ref{eq:robust.po.x}): unobserved heterogeneity matters and different values of $\sigma$ can produce different rankings over treatment policies. 

Why the difference? Criterion (\ref{eq:robust.po.x}) guards against distribution shifts where individuals may be adversarially reassigned (by shifting their $X$) into the treatment/non-treatment regions based on the size of their individual treatment effects. Criterion (\ref{eq:criterion.nohet}) does not guard against shifts with this ``selection'' feature. Instead, it guards against shifts in which individuals are reassigned into the treatment/non-treatment regions based on the size of their CATE, which is common to all individuals with the same $X$.

\subsubsection{Costly Treatment}\label{sec:cost}

The literature dealing with empirical welfare maximization (EWM) typically focuses attention on policies that attain the highest reward while ignoring cost of treatment. We have followed this convention. However, our approach easily extends to allow for a fixed treatment cost $c$ per individual, as in \cite{KitagawaTetenov2018}, by redefining $Y_1$ as the treated outcome net of that cost. Thus, a ``cost-aware'' version of the robust welfare objective (\ref{eq:robust.po.x}) is
\[
 \mr{RW}(\tau;P) =  \sup_{\eta \geq 1} \E_P \left[ \min \left\{ Y_0  + \eta h_0(X;\tau) , Y_1 - c + \eta h_1(X;\tau) \right\}   \right] - \eta \varepsilon .
\]
Similarly, cost-aware empirical robust welfare criteria can be obtained by replacing $\hat \Delta^+_i$ and $\hat \Delta_i$ in (\ref{eq:robust.po.x.empirical}) and (\ref{eq:constant}) with $\hat \Delta^+_i - c$ and $\hat \Delta_i - c$, respectively.

\section{Empirical Illustration}\label{sec:empirical}

We now present an empirical application using data from \cite{MKP2021}. The authors study the effect of the Gender and Enterprise Together business training program on various outcomes (e.g., profits) for a sample of rural Kenyan firms run by female entrepreneurs. The training focuses on both standard business topics (e.g., record keeping and pricing) and topics designed to overcome gender-related constraints (e.g., division of household and business tasks). Firm outcomes are evaluated over a four-year window following training, raising the possibility that distributions may drift over time.

The experiment used two layers of randomization: at the market level and again at the firm level. Markets were randomly assigned to treatment (at least one firm in the market is invited to training) or control (no firm is invited to training), conditional on strata. Strata were determined by geographic region and the number of firms in the market. Within treated markets, firms were randomly assigned treatment conditional on their baseline weekly profits, as reported in a pre-experiment market census. Thus, assignment is random conditional on both strata and baseline profits. 

The outcome variable $Y$ is profit one year after training.\footnote{We convert profits in years after training to real values in the year of training using Kenyan CPI data in the \cite{MKP2021} replication files.} We take the experimental population to be firms in treatment markets. We drop the few firms that do not report subsequent profits, leaving a sample of 2,009 firms, of which 1,096 were assigned the treatment. The average treatment effect is approximately 157 Kenyan Shillings (KSh), roughly 11\% of mean untreated profits.

\begin{figure}[t]
\centering
  \includegraphics[width = 0.65\textwidth]{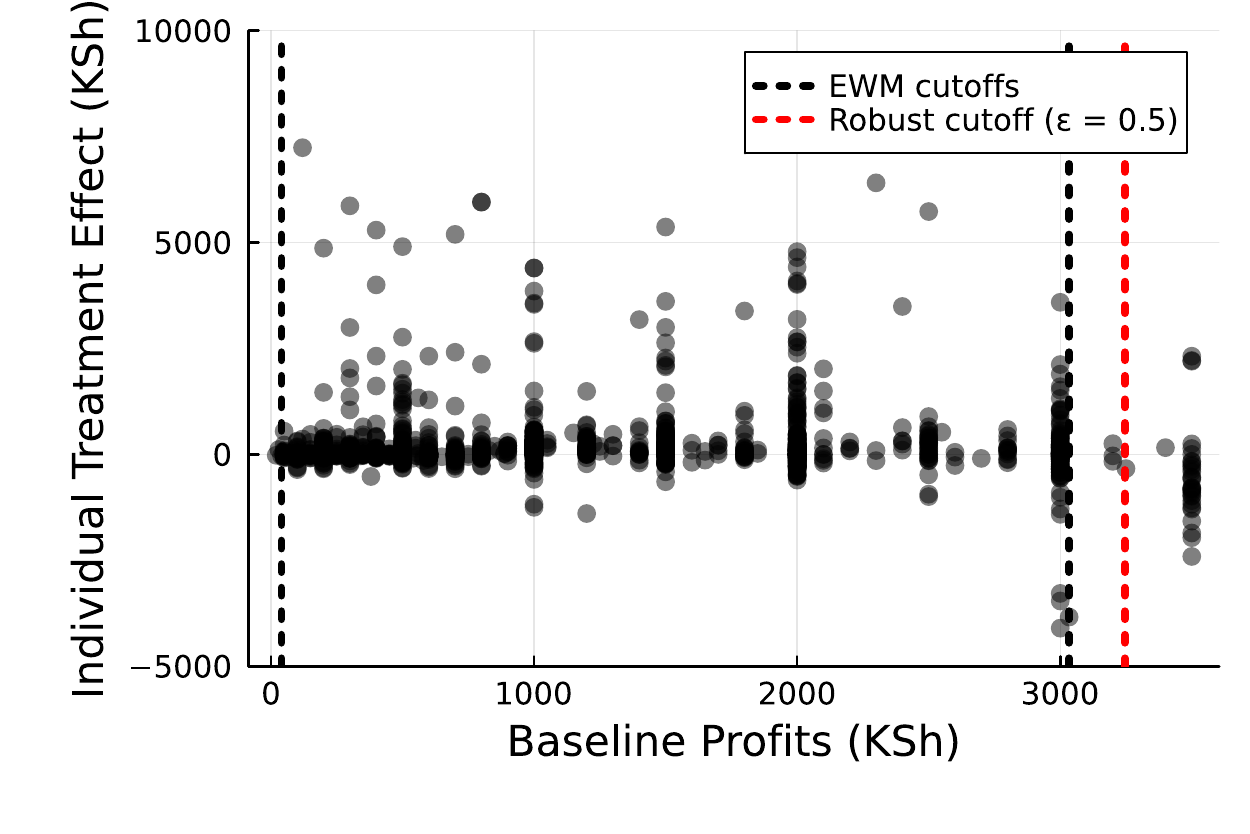}
\caption{Scatter plot of estimated individual treatment effects $\hat{\Delta}_i^+$ imputed under perfect positive dependence against baseline profits. \emph{Note:} Vertical lines indicate thresholds for the EWM and robust policies (the latter with $\varepsilon = 0.5$).}
\label{fig:mkp.scatter}
\end{figure}

We focus on a class of tree-based policies that assign treatment as a function of baseline profits (also measured in KSh). We consider trees of depth 2, so treatment is assigned depending on whether baseline profits fall in a certain interval. Within this class, the empirical welfare maximizing (EWM) policy is to treat if baseline profits exceed 40 KSh and are less than 3,034 KSh. Thus, only the least and most profitable firms are excluded from treatment under the EWM policy (see Figure~\ref{fig:mkp.scatter}). 

To examine the robustness of this policy to adversarial shifts in outcomes and baseline profits, we proceed as in Section~\ref{sec:bounds}. We impute counterfactual mappings using kernel estimates of the distribution of treated and untreated outcomes conditional on strata and baseline profits.\footnote{We estimate the conditional CDFs of treated and untreated outcomes nonparametrically using the \texttt{R} package \texttt{np} with bandwidths chosen by cross validation, treating strata as categorical. For firms reporting outcomes of zero, we draw the rank uniformly on $[0,\hat F_d(0|x)]$. Similarly, we set the imputed counterfactual to zero whenever its rank is below $\hat F_{1-d}(0|x)$.} Although we use both baseline profits and strata to impute individual causal effects, strata do not factor into our neighborhood construction. Policies are solely a function of baseline profits, so only baseline profits are relevant for computing robust welfare. We take the lower bound on the support of outcomes to be $\underline Y =0$ as the post-treatment profit distribution appears truncated at zero. Similar results are obtained with $\underline Y = -\infty$.

\begin{figure}[t]
\centering
  \includegraphics[width = 0.65\textwidth]{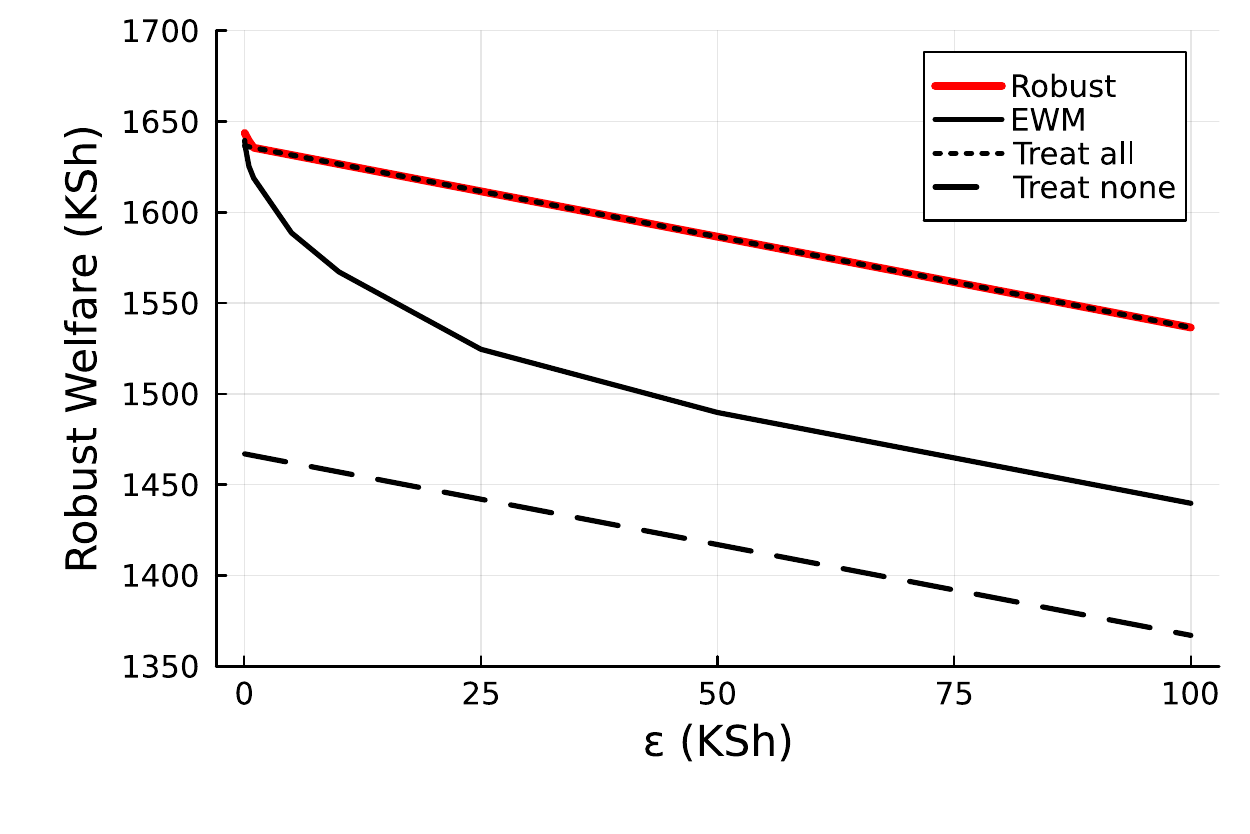}
\caption{Empirical robust welfare $\mathrm{ERW}_n(\tau;P^+)$ under perfect positive dependence. \emph{Note:} Robust corresponds to the policy that maximizes $\mathrm{ERW}_n(\tau;P^+)$ for given $\varepsilon$, EWM is the empirical welfare maximizing policy.}
\label{fig:mkp.sorting}
\end{figure}

First consider the case of perfect positive dependence. Figure~\ref{fig:mkp.sorting} plots empirical robust welfare $\mathrm{ERW}_n(\tau;P^+)$ from (\ref{eq:robust.po.x.empirical}) for the EWM policy and policies in which all firms or no firms are treated, for $\varepsilon \in \{0.05, 0.1, 0.5, 1, 5, 10, 25, 50, 75, 100\}$. For each $\varepsilon$ we also compute the robust policy $\hat \tau$ that maximizes $\mathrm{ERW}_n(\tau;P^+)$ in (\ref{eq:robust.po.x.empirical}).\footnote{To do so, we first ran a search over a fine grid of thresholds to select parameter values that approximately maximized criterion (\ref{eq:robust.po.x.empirical}). We then used these as starting values for a gradient-free optimization algorithm and selected the parameter values that yielded the highest optimum.} To interpret $\varepsilon$, note that  the estimated ATE is 157 KSh, so the neighborhood size of $\varepsilon$ includes populations with an ATE between $157-\varepsilon$ and $157+\varepsilon$ KSh. Remark~\ref{rmk:pretreat} also notes that $\varepsilon$ is the amount by which average untreated outcomes can shift between the experimental and target populations. Mean reported profits for untreated firms one and three years after treatment are approximately 1,467 KSh and 1,493 KSh, a difference of approximately 26 KSh (or 13 KSh per year). Consider a policymaker deciding whether to implement the policy four years, say, after the initial treatment. An annual drift around 13 KSh suggests that $\varepsilon \approx 50$ might be reasonable. Figure~\ref{fig:mkp.sorting} shows that even for neighborhoods of size $\varepsilon \geq 1$, the EWM policy performs much worse than the policy in which everyone is treated.

The robust policy is different for different $\varepsilon$. For very small $\varepsilon$, the robust policy treats all firms with baseline profits below a higher threshold (3,299 KSh with $\varepsilon = 0.1$ and 3,246 KSh with $\varepsilon = 0.5$) than the EWM policy. For larger values of $\varepsilon$ (say, $\varepsilon \geq 1$) the robust policy is the policy in which all firms are treated. For intuition, Figure~\ref{fig:mkp.scatter} plots the imputed individual treatment effects $\hat{\Delta}_i^+$ under perfect positive dependence against baseline profits. Under the EWM policy, all firms between the two vertical black dashed lines are treated, while those outside this interval are not treated. The upper threshold of 3,034 KSh under the EWM policy is just to the right of a large mass of firms who are positively affected (on average), but with a lot of heterogeneity. Under adversarial shifts, some of these firms who are positively affected can be reallocated across the frontier into the non-treatment region, reducing welfare. The robust policy uses a higher threshold, so that ``larger'' shifts are needed to push these firms across the frontier. For $\varepsilon = 0.5$, this threshold is roughly midway between the previously mentioned mass of firms, and another mass with higher baseline profits but who are negatively affected (on average) by treatment. As there is still a lot of heterogeneity on both sides of this threshold, for larger values of $\varepsilon$ it is optimal to choose a  policy in which all firms are treated. This finding is consistent with the numerical experiments we performed in Section~\ref{sec:po.x.implications}.

It is also striking in Figure~\ref{fig:mkp.sorting} that the robust welfare of the EWM policy decays rapidly over small neighborhoods. For instance, with $\varepsilon = 25$ the robust welfare of the EWM policy is over 100 KSh below its welfare in the experimental population. Figure~\ref{fig:mkp.scatter} shows this is explained by the large heterogeneity in individual treatment effects, so that only small shifts are required to reallocate firms near the boundary who benefit most from treatment into non-treatment regions and vice versa. Nevertheless, robust welfare of the EWM policy remains above that of the treat-none policy. Moreover, robust welfare gain of the EWM policy is positive up to a neighborhood of size 70 or so, roughly 45\% of the experimental population ATE. Thus, the EWM policy should deliver welfare improvements relative to treating no one, even allowing for a broad class of distribution shifts.

On the other hand, no individualized policy delivers better robust welfare than a one-size-fits-all policy in which all firms are treated, except for the smallest values of $\varepsilon$. And even at these small values of $\varepsilon$, the difference in $\mathrm{ERW}_n(\tau;P^+)$ for the robust policy and treat-everyone policy is less than 10 KSh, which represents only a few percent of the ATE. As 98\% of the experimental population is treated under the EWM policy, the additional cost of implementing the robust policies seems marginal relative to that of the EWM policy.


\begin{figure}[t]
\centering
  \includegraphics[width = 0.65\textwidth]{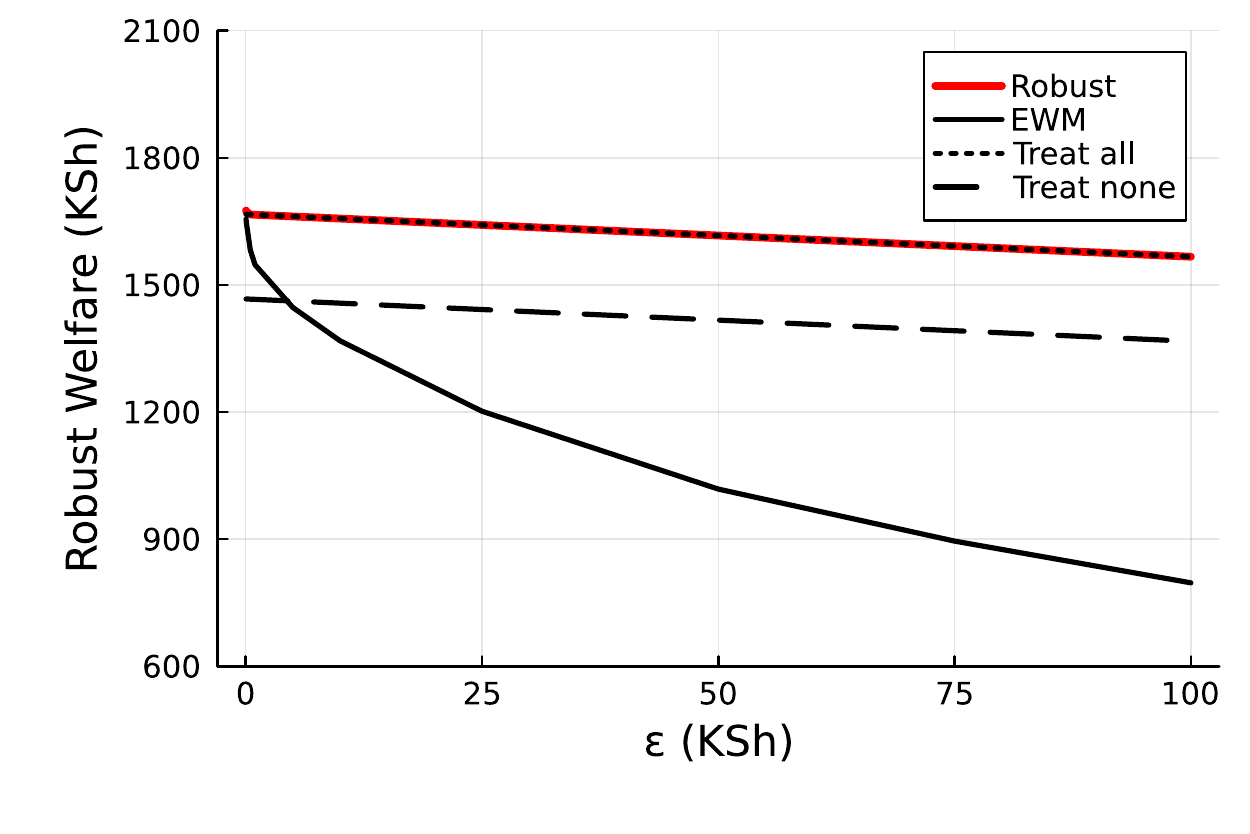}
\caption{Empirical robust welfare $\mathrm{ERW}_n(\tau;P^-)$ under perfect negative dependence. \emph{Note:} Robust corresponds to the policy that maximizes $\mathrm{ERW}_n(\tau;P^-)$ for given $\varepsilon$, EWM is the empirical welfare maximizing policy.}
\label{fig:mkp.negative}
\end{figure}

We repeat the analysis under perfect negative dependence. Results are plotted in Figure~\ref{fig:mkp.negative}. Individual treatment effects are more heterogeneous under perfect negative dependence: the standard deviations of $\hat \Delta_i^-$ and $\hat \Delta_i^+$ are approximately 3,225 and 707 KSh, respectively. In view of the discussion in Section~\ref{sec:po.x.implications}, it is not surprising that robust welfare decays much faster as $\varepsilon$ increases. Indeed, the treat-all policy maximizes $\mathrm{ERW}_n(\tau;P^-)$ for $\varepsilon \geq 0.5$. Overall, it appears preferable to adopt a policy providing training to all firms.

\section{Conclusion}

We consider the problem of learning personalized treatment policies that perform well in target populations beyond the experimental (or training) population from which the data are sampled. This paper makes two main contributions. First, we develop new methods for policy learning that are robust to shifts in the distribution of both outcomes and characteristics between the experimental and target populations. Second, we shed light on how heterogeneity in CATEs and unobserved heterogeneity in treatment effects (within the experimental population) play distinct but important roles in shaping external validity. In addition to policy learning, the methods developed in this paper may be used as a stress test to assess the fragility or robustness of treatment policies to distribution shifts.

\appendix

\section{Proofs}

\subsection{Proofs for Section~\ref{sec:neighborhood}}

\begin{proof}[Proof of Proposition~\ref{prop:ate.unbounded}]
Follows from Proposition~\ref{prop:ate} with $\ul Y = -\infty$ and $\ol Y = +\infty$.
\end{proof}

\subsection{Proofs for Section~\ref{sec:po}}

\begin{proof}[Proof of Proposition~\ref{prop:po.unbounded}]
Follows from Proposition~\ref{prop:po} with $\ul Y = -\infty$.
\end{proof}

\bigskip

\begin{proof}[Proof of Proposition~\ref{prop:po.gain.unbounded}]
Follows from Proposition~\ref{prop:po.gain} with $\ul Y = -\infty$ and $\ol Y = +\infty$.
\end{proof}

\subsection{Proofs for Section~\ref{sec:po.x}}

\begin{proof}[Proof of Proposition~\ref{prop:po.x.unbounded}]
Follows from Propositions~\ref{prop:po.x} and~\ref{prop:po.gain.x} with $\ul Y = -\infty$ and $\ol Y = +\infty$.
\end{proof}

\bigskip

We first state two general consistency and convergence rate results for the criterion
\[
 \mr{ERW}_n(\tau;P) =  \max\left\{ \bar Y_0 +  \sup_{\eta \geq 1} \frac{1}{n} \sum_{i=1}^n \min\left\{  \hat \Delta_i + \eta h_1(X_i; \tau), \eta h_0(X_i; \tau) \right\}  - \eta\varepsilon \,,\,\underline Y \right\} ,
\]
with $P$ a generic experimental population, $\hat \Delta_i$ denoting an estimate of individual $i$'s treatment effect $\Delta_i$ under $P$, based on a random sample of size $n$ from the experimental population, and $\ul Y := \inf \mc Y$. Propositions~\ref{prop:robust.po.x.empirical.bounds}-\ref{prop:robust.po.x.empirical.1.rate} follow from these results.

\begin{proposition}\label{prop:robust.po.x.consistent}
Suppose that the conditions of Proposition~\ref{prop:po.x} hold, $\mc X$ is a bounded subset of $\mb R^d$, $\frac{1}{n} \sum_{i=1}^n  |\hat \Delta_i - \Delta_i| \to_p 0$, and $\bar Y_0 - \E_P[Y_0] \to_p 0$. Then
\begin{enumerate}[nosep, left=0pt]
\item $\sup_{\tau \in \mc T} |\mr{ERW}_n(\tau;P) - \mr{RW}(\tau;P)| \to_p 0$; 
\item $\sup_{\tau \in \mc T}\mr{RW}(\tau;P) - \mr{RW}(\hat \tau;P) \to_p 0$ for any maximizer $\hat \tau$ of $\mr{ERW}_n(\,\cdot\,;P)$.
\end{enumerate}
Moreover, the convergence in parts~1. and~2.~holds uniformly for $\varepsilon \geq \ul \varepsilon$ for any arbitrarily small $\ul \varepsilon > 0$.
\end{proposition}

\begin{proposition}\label{prop:robust.po.x.rate}
Suppose that the conditions of Proposition~\ref{prop:po.x} hold,  $\E_P[Y_d^2] < \infty$ for $d = 0,1$, $\mc C^*$ is a VC class of dimension $v_n \gtrsim 1$, and there are positive constants $a_n, b_n$ such that $\frac{1}{n} \sum_{i=1}^n |\hat \Delta_i - \Delta_i| = O_p(a_n)$ and $|\bar Y_0 - \E_P[Y_0]| = O_p(b_n)$. Then with $c_n = \max\{a_n, b_n, (v_n / n)^{1/2}\}$,
\begin{enumerate}[nosep, left=0pt]
\item $\sup_{\tau \in \mc T} |\mr{ERW}_n(\tau;P) - \mr{RW}(\tau;P)| = O_p( c_n)$; 
\item $\sup_{\tau \in \mc T}\mr{RW}(\tau;P) - \mr{RW}(\hat \tau;P) = O_p( c_n)$ for any maximizer $\hat \tau$ of $\mr{ERW}_n(\,\cdot\,;P)$.
\end{enumerate}
Moreover, the convergence rates in parts~1. and~2.~hold uniformly for $\varepsilon > 0$.
\end{proposition}

\begin{proof}[Proof of Propositions~\ref{prop:robust.po.x.empirical.bounds} and \ref{prop:robust.po.x.empirical.bounds.rate}]
Immediate from Propositions~\ref{prop:robust.po.x.consistent} and~\ref{prop:robust.po.x.rate}, setting $P = P^{\pm}$ and $\Delta = \Delta^\pm$.
\end{proof}

\bigskip

\begin{proof}[Proof of Propositions~\ref{prop:robust.po.x.empirical.1} and \ref{prop:robust.po.x.empirical.1.rate}]
Immediate from Propositions~\ref{prop:robust.po.x.consistent} and~\ref{prop:robust.po.x.rate}, invoking (\ref{eq:po.degenerate}) and setting $\Delta = \delta(X)$.
\end{proof}

\bigskip

We first give a preliminary result before giving the proof of Proposition~\ref{prop:robust.po.x.consistent}.

\begin{lemma}\label{lem:truncate.eta}
Fix $\ul \varepsilon > 0$ and let $C = 1 + 2 \ul \varepsilon^{-1} \E_P[|\Delta|]$. Then 
\begin{multline*}
 \sup_{\eta \geq 1} \E_P \left[ \min\left\{  \Delta + \eta h_1(X; \tau), \eta h_0(X; \tau) \right\} \right] - \eta \varepsilon \\
  =  \sup_{1 \leq \eta \leq C} \E_P \left[ \min\left\{  \Delta + \eta h_1(X; \tau), \eta h_0(X; \tau) \right\} \right] - \eta \varepsilon
\end{multline*}
for all $\tau : \mc X \to \{0,1\}$ and $\varepsilon \geq \ul \varepsilon$.
\end{lemma}

\begin{proof}[Proof of Lemma \ref{lem:truncate.eta}]
First note by non-negativity of $\eta$, $h_0$, and $h_1$ and the fact that for each $x$ at least one of $h_0(x;\tau)$ and $h_1(x;\tau)$ is zero, we have
\begin{equation} \label{eq:truncate.eta.1}
 |\min\{  \Delta + \eta h_1(x; \tau), \eta h_0(x; \tau) \}| \leq |\Delta|\,.
\end{equation}
Take any $\tau$ and any $\eta \geq 1$ and $\varepsilon \geq \ul \varepsilon$. Then 
\begin{multline*}
 \E_P \left[ \min\{  \Delta + \eta h_1(X; \tau), \eta h_0(X; \tau) \} \right] - \eta \varepsilon \geq \E_P \left[ \min\{  \Delta + h_1(X; \tau), h_0(X; \tau) \} \right] - \varepsilon \\
 \iff \E_P \left[ \min\{  \Delta + \eta h_1(X; \tau), \eta h_0(X; \tau) \} - \min\{  \Delta + h_1(X; \tau), h_0(X; \tau) \} \right] \geq (\eta - 1) \varepsilon.
\end{multline*}
Then by (\ref{eq:truncate.eta.1}),
\[
 2 \E_P[|\Delta|] \geq (\eta - 1) \varepsilon \geq (\eta - 1) \ul \varepsilon.
\]
Hence, an optimal $\eta$ will never exceed $1 + 2 \ul \varepsilon^{-1} \E_P[|\Delta|]$.
\end{proof}

\bigskip

\begin{proof}[Proof of Proposition~\ref{prop:robust.po.x.consistent}]
It suffices to prove part 1., as part 2. then follows easily.
By Lemma~\ref{lem:truncate.eta} and the condition $\frac{1}{n} \sum_{i=1}^n |\hat \Delta_i - \Delta_i| \to_p 0$, wpa1 the optimal $\eta$ is in the interval $[1, 2 + 2 \ul \varepsilon^{-1} \E_P[|\Delta|]] =: [1, C]$ for both $\mr{RW}(\tau;P)$ and  $\mr{ERW}_n(\tau;P)$, uniformly for $\tau \in \mc T$ and for $\varepsilon \geq \ul \varepsilon$. It then follows from the fact that the $\min$ and $\max$ operations are Lipschitz that wpa1, 
\begin{multline}\label{eq:robust.po.x.empirical.bounds.1}
 |\mr{ERW}_n(\tau;P) - \mr{RW}(\tau;P)| \leq | \bar Y_0 - \E_P[Y_0]| + \frac{1}{n} \sum_{i=1}^n |\hat \Delta_i - \Delta_i| \\
 + \sup_{\eta \in [1,C]} \Bigg| \frac{1}{n} \sum_{i=1}^n \min\{   \Delta_i + \eta h_1(X_i; \tau), \eta h_0(X_i; \tau) \} \\
  - \E_P \left[ \min\{ \Delta + \eta h_1(X; \tau), \eta h_0(X; \tau) \} \right] \Bigg|
\end{multline}
holds uniformly for $\tau \in \mc T$ and $\varepsilon \geq \ul \varepsilon$. The first two terms on the right-hand side of (\ref{eq:robust.po.x.empirical.bounds.1}) are $o_p(1)$ by assumption. If $\tau \equiv 0$ then $\min\{   \Delta + \eta h_1(X; \tau), \eta h_0(X; \tau) \} = 0$ and the third term in (\ref{eq:robust.po.x.empirical.bounds.1}) disappears. If $\tau \equiv 1$ then $\min\{   \Delta + \eta h_1(X; \tau), \eta h_0(X; \tau) \} = \Delta$ and the third term in (\ref{eq:robust.po.x.empirical.bounds.1}) is $o_p(1)$ by the weak law of large numbers. For any $\tau$ that is not identically $0$ or $1$, we have $|h_d(x;\tau)-h_d(\tilde x;\tau)| \leq \|x - \tilde x\|$, for $d = 0,1$. Therefore, $\{\eta h_d : \tau \in \mc T, \eta \in [1,C], d = 0,1\}$ is a set of Lipschitz functions on $\mc X$ with Lipschitz constant at most $C$, which are uniformly bounded by $C \times \mathrm{diam}(\mc X)$. As $\mc X$ is bounded, this class has finite $L^1(P)$ bracketing numbers \cite[Corollary 2.7.2]{vanderVaartWellner}. It follows that $\{ \min\left\{  \Delta + \eta h_1(X;\tau), \eta h_0(X;\tau) \right\}  : \tau \in \mc T, \eta \in [1,C] \}$ also has finite $L^1(P)$ bracketing numbers. The third term in (\ref{eq:robust.po.x.empirical.bounds.1}) is therefore $o_p(1)$ uniformly for $\tau \in \mc T$ and $\varepsilon \geq \ul \varepsilon$ by the Glivenko--Cantelli theorem \cite[Theorem 2.4.1]{vanderVaartWellner}.
\end{proof}

\bigskip

We first give a preliminary result before giving the proof of Proposition~\ref{prop:robust.po.x.rate}.

\begin{lemma}\label{lem:vc}
Suppose $\mc C^*$ is a VC class with dimension $v$. Then the class of functions
\[
 \left\{(\Delta,x) \mapsto \min\{ \Delta + \eta h_1(x;\tau) , \eta h_0(x;\tau) \}: \eta \geq 1, \tau \in \mc T\right\}
\]
is VC subgraph with dimension at most $2v$.
\end{lemma}

\begin{proof}[Proof of Lemma~\ref{lem:vc}]
We first show that $\{x \mapsto \eta h_d(x;\tau) : \eta \geq 1, \tau \in \mc T, d = 0,1\}$ is VC subgraph.
That is, we need to show that the set of all subgraphs of the form $\{(x,t) : t < \eta \inf_{\tilde x \in C} \|x - \tilde x\|\}$ for $\eta \geq 1$ and $C \in \mc C \cup \mc C^c$ forms a VC class of sets in $\mc X \times \mb R$. It is without loss of generality to consider $t \geq 0$ since $\eta \inf_{\tilde x \in C} \|x - \tilde x\| \geq 0$. Note that
\[
 t < \eta \inf_{\tilde x \in C} \|x - \tilde x\| \iff t/\eta < \inf_{\tilde x \in C}\|x - \tilde x\| \iff (x,t) \not \in \left\{(x,s) : \inf_{\tilde x \in C} \|x - \tilde x\| \leq \frac{s}{\eta} \right\} .
\]
Taking complements preserves VC dimension, so it suffices that $\mc C^*$ is a VC class.

To complete the proof, note that $h_0(x;\tau),h_1(x;\tau) \geq 0$ and $h_0(x;\tau) h_1(x;\tau) = 0$. Hence, 
\[
 t < \min\{ \Delta + \eta h_1(x;\tau) , \eta h_0(x;\tau) \}
 \iff
 \begin{cases}
 t < \Delta + \eta h_1(x;\tau), & \text{if } t < 0, \\
 t < \min\{\Delta , \eta h_0(x;\tau)\}, & \text{if } t \geq 0.
 \end{cases}
\]
By Lemmas 2.6.17 and 2.6.18 of \cite{vanderVaartWellner} and our above argument, $\{\{(\Delta,x,t) : t < \Delta + \eta h_1(x;\tau)\}: \eta \geq 1, \tau \in \mathcal T \}$ and $\{\{(\Delta,x,t) : t < \min\{\Delta , \eta h_0(x;\tau)\}\}: \eta \geq 1, \tau \in \mathcal T \}$ are both VC with dimension at most $v$. Combining cases, the total dimension is at most $2v$.
\end{proof}

\bigskip

\begin{proof}[Proof of Proposition~\ref{prop:robust.po.x.rate}]
It suffices to prove part 1., as part 2. then follows easily.
In view of (\ref{eq:robust.po.x.empirical.bounds.1}), we only need to show that
\begin{multline}\label{prop:robust.po.x.empirical.bounds.rate.1}
\sup_{\tau \in \mc T, \eta \geq 1} \Bigg| \frac{1}{n} \sum_{i=1}^n \min\{   \Delta_i + \eta h_1(X_i; \tau), \eta h_0(X_i; \tau) \} \\
  - \E_P \left[ \min\{ \Delta + \eta h_1(X; \tau), \eta h_0(X; \tau) \} \right] \Bigg| = O_p((v_n/n)^{1/2}).
\end{multline}
Note also that we do not need to restrict $\eta$, as by our assumption on $\mc C^*$ and Lemma~\ref{lem:vc}, 
\[
 \mc F :=  \left\{(\Delta,x) \mapsto \min\{ \Delta + \eta h_1(x;\tau) , \eta h_0(x;\tau) \}: \eta \geq 1, \tau \in \mc T\right\}
\]
is VC subgraph with dimension at most $2v_n$. Note by non-negativity of $h_0$ and $h_1$ and the fact that $h_0(x;\tau) h_1(x;\tau) = 0$ for all $x, \tau$, we have that $|f| \leq |\Delta|$ for each $f \in \mc F$. Hence $F := |\Delta|$ is an envelope of $\mc F$. By Theorem 2.6.7 of \cite{vanderVaartWellner}, there is a universal constant $C \geq 1$ such that for any probability measure $\mu$ and $t \in (0,1)$,
\[
 N(\mc F, L^2(\mu), t\|F\|_{L^2(\mu)}) \leq C (2v_n + 1) (16e)^{2v_n + 1} t^{-4v_n}.
\]
Hence there is a universal constant $C'$ such that for $0 < \delta \leq 1$,
\[
 J(\mc F, F, \delta) := \int_0^\delta \sup_{\mu} \sqrt{ \log  2N(\mc F, L^2(\mu), t\|F\|_{L^2(\mu)}) } \, d t \leq C' \sqrt{v_n}\,,
\]
where the supremum is taken over all discrete probability measures with a finite number of atoms. Let $\| \nu_n\|_{\mc F}$ denote the left-hand side of (\ref{prop:robust.po.x.empirical.bounds.rate.1}) multiplied by $n^{1/2}$. By Remark 3.5.5 of \cite{GineNickl2016} we have the bound 
\[
 \E_P[ \| \nu_n\|_{\mc F} ] \leq C' 8 \sqrt{ 2 v_n} (\E_P[\Delta^2])^{1/2}.
\]
For any coupling of $Y_0|X$ and $Y_1|X$, we have the trivial bound $|\Delta| \leq |Y_0| + |Y_1|$, 
so the result follows by Markov's inequality and the finite second moment of $Y_0,Y_1$.
\end{proof}

\bigskip

\begin{proof}[Proof of Proposition~\ref{prop:robust.po.x.empirical.3}]
It suffices to prove part 1., as part 2. then follows easily. To simplify notation, let $\hat G^{(i)}_{\eta,\tau}(z) = \hat G_{\eta,\tau}(z|X_i,V_i) $, and similarly for $G_{\eta,\tau}$, $\hat F_0$, $\hat F_1$, $F_0$, and $F_1$. By the functional form of $\mr{ERW}_n$, it suffices to show
\begin{multline*}
 \sup_{\tau \in \mc T} \bigg| \sup_{\eta \geq 1}\bigg( \frac{1}{n} \sum_{i=1}^n \int z \, \mr d \hat G_{\eta,\tau}^{(i)}(z)  - \eta\varepsilon \bigg) \\
 - \bigg(  \sup_{\eta \geq 1} \int \int z \, \mr d G_{\eta,\tau}(z|x,v) \, \mr d P_{X,V}(x,v) - \eta \varepsilon \bigg) \bigg| = o_p(1).
\end{multline*}
We first show that there is a sufficiently large constant $B$ such that for all $\tau \in \mc T$, the argsup of both problems is an element of $[1,B]$ wpa1. For the first problem, suppose the assertion is false. Then for some $\tau \in \mc T$, $\eta \geq B$, and $\varepsilon \geq \ul \varepsilon$, we have
\begin{equation}\label{eq:prop:robust.po.x.empirical.3}
 \frac{1}{n} \sum_{i=1}^n \int z \, \mr d ( \hat G_{\eta,\tau}^{(i)}(z) - \hat G_{1,\tau}^{(i)}(z)) \geq (\eta - 1) \varepsilon \geq (B - 1) \varepsilon \geq (B - 1) \ul \varepsilon.
\end{equation}
Fix $1 \leq i \leq n$ and suppose that $h_0(X_i;\tau) = 0$. Then 
\begin{align*}
 & \left| \int z \, \mr d ( \hat G_{\eta,\tau}^{(i)}(z) - \hat G_{1,\tau}^{(i)}(z)) \right| \\
 & = \left| \int z \, \mr d \left( \left(\hat F_1^{(i)}(z - \eta h_1(X_i;\tau)) - \hat F_1^{(i)}(z - h_1(X_i;\tau))  \right) \left(1 - \hat F_0^{(i)}(z) \right) \right) \right|  \\
 & \leq  \int |z| \, \mr d \hat F_0^{(i)}(z) + \int |z| \, \mr d \hat F_1^{(i)}(z) .
\end{align*}
A symmetric argument applies when $h_1(X_i;\tau) = 0$. Hence, the left-hand side of (\ref{eq:prop:robust.po.x.empirical.3}) is bounded above by
\[
 \int |z| \, \mr d \bar F_0(z)  + \int |z| \, \mr d \bar F_1(z)  .
\]
It follows from the integrability condition in the statement of the proposition that $B$ can be chosen sufficiently large so that inequality (\ref{eq:prop:robust.po.x.empirical.3})  is violated wpa1. That the argsup of the second problem is an element of $[1,B]$ may be deduced similarly by contradiction. It therefore suffices to show
\[
 \sup_{\tau \in \mc T, \eta \in [1,B]} \bigg| \frac{1}{n} \sum_{i=1}^n \int z \, \mr d \hat G_{\eta,\tau}^{(i)}(z) 
 - \int \int z \, \mr d G_{\eta,\tau}(z|x,v) \, \mr d P_{X,V}(x,v)  \bigg| = o_p(1).
\]

By similar arguments to the above, we may use the integrability condition in the statement of the proposition to deduce that for any $\delta > 0$ there exists a finite constant $M$ such that wpa1 the inequalities
\[
 \frac{1}{n} \sum_{i=1}^n \int_{|z| > M } |z| \, \mr d \hat G_{\eta,\tau}(z|X_i,V_i)  \leq \delta \,, \quad \int \int_{|z|>M} |z| \, \mr d G_{\eta,\tau}(z|x,v) \, \mr d P_{X,V}(x,v) \leq \delta
\]
hold uniformly in $(\eta,\tau) \in [1,B] \times \mc T$. Then letting $(z)_M = \max\{-M,\min\{z,M\}\}$ and using the fact that $|z - (z)_M| \leq (|z| - M) \mb I[|z| > M] \leq |z| \mb I[|z| > M]$, we have
\begin{multline*}
 \Bigg| \frac{1}{n} \sum_{i=1}^n \int z \, \mr d \hat G_{\eta,\tau}^{(i)}(z) 
 - \int \int z \, \mr d G_{\eta,\tau}(z|x,v) \, \mr d P_{X,V}(x,v) \\
 - \Bigg( \frac{1}{n} \sum_{i=1}^n \int (z)_M \, \mr d \hat G_{\eta,\tau}^{(i)}(z) 
 - \int \int (z)_M \, \mr d G_{\eta,\tau}(z|x,v) \, \mr d P_{X,V}(x,v) \Bigg) \Bigg|  \leq 2\delta,
\end{multline*}
wpa1. Finally, it remains to show
\[
 \sup_{\tau \in \mc T, \eta \in [1,B]} \bigg| \frac{1}{n} \sum_{i=1}^n \int (z)_M \, \mr d \hat G_{\eta,\tau}^{(i)}(z) 
 - \int \int (z)_M \, \mr d G_{\eta,\tau}(z|x,v) \, \mr d P_{X,V}(x,v)  \bigg| = o_p(1).
\]
Now write
\[
\begin{aligned}
 & \bigg| \frac{1}{n} \sum_{i=1}^n \int (z)_M \, \mr d \hat G_{\eta,\tau}^{(i)}(z) 
 - \int \int (z)_M \, \mr d G_{\eta,\tau}(z|x,v) \, \mr d P_{X,V}(x,v)  \bigg| \\
 & \leq \left| \frac{1}{n} \sum_{i=1}^n \int (z)_M \, \mr d \hat G_{\eta,\tau}^{(i)}(z) - \frac{1}{n} \sum_{i=1}^n \int (z)_M \, \mr d G_{\eta,\tau}^{(i)}(z)\right| \\
 & \quad + \left| \frac{1}{n} \sum_{i=1}^n \int (z)_M \, \mr d G_{\eta,\tau}^{(i)}(z) - \int \int (z)_M \, \mr d G_{\eta,\tau}(z|x,v) \, \mr d P_{X,V}(x,v)\right| =: T_1 + T_2. 
\end{aligned}
\]
For any CDF $G$, by integration by parts, we have $\int (z)_M \, \mr d G(z) = M - \int_{-M}^M G(t) \, \mr d t$. Hence for $T_1$, we have
\begin{multline*}
 \left| \int (z)_M \, \mr d \left(\hat G_{\eta,\tau}^{(i)}(z) - G_{\eta,\tau}^{(i)}(z) \right) \right| 
 \leq 2 M \sup_z | \hat G_{\eta,\tau}^{(i)}(z) - G_{\eta,\tau}^{(i)}(z)| \\
 \leq 8M \max_{d = 0,1} \sup_y \left|\hat F_d^{(i)}(y) - F_d^{(i)}(y) \right|. 
\end{multline*}
It follows by the uniform consistency condition for $\hat F_0$ and $\hat F_1$ in the statement of the proposition that $\sup_{\tau \in \mc T, \eta \in [1,B]} |T_1| \to_p 0$. To establish the corresponding result for $T_2$, it suffices to show that a uniform law of large numbers holds for the class of functions
\[
 \left\{ (x,v) \mapsto \E[(a(X,Y_0,Y_1))_M | X = x,V = v] : a \in \mc A \right\},
\]
where $\mc A =  \left\{ (x,y_0,y_1) \mapsto \min\left\{  y_0 + \eta h_0(x;\tau)\,, y_1 + \eta h_1(x;\tau) \right\}  : \tau \in \mc T, \eta \in [1,B] \right\}$. This may be deduced by similar arguments to the proof of Proposition~\ref{prop:robust.po.x.consistent}, using the fact that $z \mapsto (z)_M$ is monotone and Lipschitz.
\end{proof}

{

\let\oldbibliography\thebibliography
\renewcommand{\thebibliography}[1]{\oldbibliography{#1}
\setlength{\itemsep}{0pt}}

\putbib
}
\end{bibunit}

\newpage

\section*{Supplemental Appendix}
\setcounter{page}{1}

\begin{bibunit}

This supplemental appendix presents some additional results extending those from the main text. Appendix~\ref{ax:bounded} generalizes the robust welfare calculations to the case of bounded potential outcomes. Appendix~\ref{ax:extensions} discusses extensions of Proposition~\ref{prop:po.unbounded} to other Wasserstein metrics. Appendix~\ref{sec:shift.x.known} extends results from Section~\ref{sec:po} to cases where there is a known shift in the distribution of characteristics. All proofs are presented in Appendix~\ref{ax:proofs.2}.

\section{Bounded Potential Outcomes}\label{ax:bounded}

In the main text we presented results for the case in which potential outcomes are supported on $\mathbb R$. Here we  generalize the results from the main text to allow for the support $\mathcal Y$ to be bounded, in the sense that at least one of  $\underline Y := \inf \mc Y$ and $\overline Y := \sup \mc Y$ is finite. For instance, $\mc Y = \{0,1\}$, $\underline Y = 0$, and $\overline Y =1$ for binary outcomes, while $\mc Y = \mb R_+$, $\underline Y = 0$, and $\overline Y = +\infty$ for non-negative outcomes (e.g., earnings).  

\subsection{Section~\ref{sec:neighborhood}}

Proposition~\ref{prop:ate.unbounded} is a special case of the following result with $\ul Y = -\infty$ and $\ol Y = +\infty$.

\begin{proposition}\label{prop:ate}
Suppose that $\mc Q$ is defined using $d_W(P,Q)$ induced by (\ref{eq:metric.general}). Then
\[
 \begin{aligned}
 \inf_{Q \in \mc Q} \E_Q\left[ \Delta \right] 
 & = \max \left\{ 
 \E_P \left[ \Delta \right]  - \varepsilon \,,\,
 \underline Y - \overline Y \right\}, \\
 \sup_{Q \in \mc Q} \E_Q\left[ \Delta \right] 
 & = \min \left\{ 
 \E_P \left[ \Delta \right]  + \varepsilon \,,\,
 \overline Y - \underline Y \right\}.
 \end{aligned}
\]
\end{proposition}

\noindent
Note that if either $\ul Y = -\infty$ or $\ol Y = + \infty$, then (\ref{eq:bounds.ate}) holds and $\varepsilon$ is the maximum that the ATE can differ between the experimental and target populations.

\subsection{Section~\ref{sec:po}}

Propositions~\ref{prop:po.unbounded} and~\ref{prop:po.gain.unbounded} are special cases of the following with $\ul Y = -\infty$ and $\ol Y = +\infty$. 
 
\begin{proposition}\label{prop:po}
Suppose that $\mc Q$ is defined using the Wasserstein metric $d_W(P,Q)$ induced by (\ref{eq:metric.po}). Then for any policy $\tau$,
\[
 \mr{RW}(\tau;P) = \max \Big\{ 
 \mr{W}(\tau; P) - \varepsilon \,,\,
 \underline Y \Big\}.
\]
\end{proposition}

\begin{proposition}\label{prop:po.gain}
Suppose that $\mc Q$ is defined using the Wasserstein metric $d_W(P,Q)$ induced by (\ref{eq:metric.po}). Then for any policy $\tau$,
\[
 \mr{RWG}(\tau;P) = \max \Big\{ 
 \mr{WG}(\tau; P) - \varepsilon \,,\,
 \E_P[\tau(X)](\underline Y - \overline Y) \Big\}.
\]
\end{proposition}

\begin{remark} \normalfont
It follows from Proposition~\ref{prop:po} that the regret of any estimated policy $\hat \tau$ under criterion (\ref{eq:robust.welfare.criterion}) is bounded by its regret under  criterion (\ref{eq:social.welfare.criterion}):
\[
 \sup_{\tau \in \mc T} \mr{RW}(\tau;P) - \mr{RW}(\hat \tau;P) \leq \sup_{\tau \in \mc T} \mr W(\tau;P) - \mr W(\hat \tau;P) \quad \mbox{for all $\varepsilon > 0$.}
\]
Hence, policy learning methods with good (statistical) regret guarantees under criterion (\ref{eq:social.welfare.criterion}) also enjoy good regret guarantees under its robust counterpart (\ref{eq:robust.welfare.criterion}).
\end{remark}

\begin{remark}\label{rmk:support} \normalfont
When $\mc Y$ is a strict subset of $\mb R$, there may be many distributions $Q \in \mc Q$ under which the support of potential outcomes is different from $\mc Y$. This raises the concern that $\mc Q$ is ``too large'', in the sense that it contains distributions with supports that the analyst would never confront in any realistic target population. The proof of Proposition~\ref{prop:po} shows that if $\underline Y = -\infty$ or $\min \mc Y = \underline Y > -\infty$ holds, then the \emph{worst-case} distributions that solve the minimization problem (\ref{eq:robust.welfare.criterion}) have support $\mc Y$. An analogous result holds for welfare gain. The neighborhoods $\mc Q$ are therefore not too large, because the worst-case distributions that are being guarded against are precisely those with support $\mc Y$.
\end{remark}

\subsection{Section~\ref{sec:po.x}}

Proposition~\ref{prop:po.x.unbounded} is a special case of the following two propositions with $\ul Y = -\infty$ and $\ol Y = +\infty$. Recall the functions $h_0$ and $h_1$ defined in Proposition~\ref{prop:po.x.unbounded}.

\begin{proposition}\label{prop:po.x}
Suppose that $\mc Q$ is defined using $d_W(P,Q)$ induced by (\ref{eq:metric.po.x}) and that $\E_P[\|X\|]$ is finite. Then for any policy $\tau$, 
\[
 \mr{RW}(\tau;P) = \max \left\{ \sup_{\eta \geq 1} \E_P \left[ \min \left\{ Y_0  + \eta h_0(X;\tau) , Y_1 + \eta h_1(X;\tau) \right\}   \right] - \eta \varepsilon \,,\, \underline Y \right\} .
\]
\end{proposition}

\begin{proposition}\label{prop:po.gain.x}
Suppose that $\mc Q$ is defined using $d_W(P,Q)$ induced by (\ref{eq:metric.po.x}) and that $\E_P[\|X\|]$ is finite. Then for any policy $\tau \not \equiv 0$, 
\begin{multline*}
 \mr{RWG}(\tau;P) = \max \Bigg\{ \sup_{\eta \geq 1} \E_P \left[ \min \left\{ \eta h_0(X;\tau) , Y_1 - Y_0 + \eta h_1(X;\tau) \right\}   \right] - \eta \varepsilon \,, \\
 \sup_{\eta \in [0,1)} \E_P \left[ \min \left\{ \eta h_0(X;\tau) , (1-\eta) (\ul Y - \ol Y) + \eta(Y_1 - Y_0) + \eta h_1(X;\tau) \right\}   \right] - \eta \varepsilon \Bigg\} .
\end{multline*}
If $\tau \equiv 0$, then $\mr{RWG}(\tau;P) = \mr{WG}(\tau;P) = 0$.
\end{proposition}

With bounded outcomes, the empirical robust welfare criterion in (\ref{eq:robust.po.x.empirical})  becomes 
\[
 \mr{ERW}_n(\tau;P^+)  = \max\left\{ \bar Y_0 +  \sup_{\eta \geq 1} \frac{1}{n} \sum_{i=1}^n \min\left\{  \hat \Delta_i^{\!+} + \eta h_1(X_i; \tau), \eta h_0(X_i; \tau) \right\}  - \eta\varepsilon \,,\,\underline Y \right\} .
\]
Propositions~\ref{prop:robust.po.x.empirical.bounds} and~\ref{prop:robust.po.x.empirical.bounds.rate} are proved for this more general criterion. Similarly, criterion (\ref{eq:constant}) becomes
\[
 \mr{ERW}_n(\tau;P)  = \max\left\{ \bar Y_0 +  \sup_{\eta \geq 1} \frac{1}{n} \sum_{i=1}^n \min\left\{  \hat \Delta_i + \eta h_1(X_i; \tau), \eta h_0(X_i; \tau) \right\}  - \eta\varepsilon \,,\,\underline Y \right\} .
\]
Propositions~\ref{prop:robust.po.x.empirical.1} and \ref{prop:robust.po.x.empirical.1.rate} are proved for this more general criterion. Finally, criterion (\ref{eq:kernel}) becomes
\[
 \mr{ERW}_n(\tau;P) = \max\left\{  \sup_{\eta \geq 1} \frac{1}{n} \sum_{i=1}^n \int z \, \mr d \hat G_{\eta,\tau}(z|X_i,V_i)  - \eta\varepsilon \,,\,\underline Y \right\} .
\]
Proposition~\ref{prop:robust.po.x.empirical.3} is proved for this more general criterion.

\section{Other Wasserstein Metrics}\label{ax:extensions}

Here we show that the conclusion of Proposition~\ref{prop:po.unbounded} is not specific to our choice of Wasserstein metric of order $1$. Let $\mc Q = \{Q : d_{W,p}(P,Q) \leq \varepsilon\}$ where $d_{W,p}(P,Q)$ is the Wasserstein metric of order $p$ for $1 \leq p < \infty$ induced by 
\begin{equation}\label{eq:metric.po.Wp}
 d((x,y_0,y_1),(\tilde x, \tilde y_0, \tilde y_1)) = \left( |y_0 - \tilde y_0|^p + |y_1 - \tilde y_1|^p + \infty \times \mathbb I[x \neq \tilde x] \right)^{1/p}.
\end{equation}
Let $\mc Y = \mathbb R$. We have the following version of Proposition~\ref{prop:po.unbounded}:

\begin{proposition}\label{prop:po.Wp}
Suppose that $\mc Q$ is defined using the Wasserstein metric $d_{W,p}(P,Q)$ induced by (\ref{eq:metric.po.Wp}) and that $Y_0$ and $Y_1$ have finite $p$th moments under $P$. Then for any policy $\tau$, 
\[
 \mr{RW}(\tau;P) = \mr{W}(\tau; P)  - \varepsilon.
\]
\end{proposition}

Together, Propositions~\ref{prop:po.unbounded} and~\ref{prop:po.Wp} provide a stronger sense in which policies with good guarantees under criterion (\ref{eq:social.welfare.criterion}) also have good external validity guarantees with respect to shifts in potential outcomes.

\section{Known Shifts in Characteristics}\label{sec:shift.x.known}

Here we extend results from Section~\ref{sec:po} to allow for known shifts in characteristics. 
In some cases, the analyst may be able to estimate (or have prior knowledge of) the distribution of characteristics in the target population. For instance, an experiment may sample one region but the analyst wishes to choose a policy for several neighboring regions. The distribution of characteristics in neighboring regions may be known, e.g., from census or administrative data, and we might reasonably assume that the distribution of potential outcomes is similar, though not the same, across regions. 

 In this case, one could consider the re-weighted social welfare criterion 
\begin{equation}\label{eq:social.welfare.reweight}
 \mathrm{W}_\rho(\tau;P) = \E_P\left[ \left( Y_1 \tau(X) + Y_0 (1-\tau(X)) \right) \rho(X) \right]
\end{equation}
where $\rho(x) = q(x)/p(x)$ is the ratio of the marginal densities for $X$ in the target and experimental populations, respectively. This criterion was considered by \cite{KitagawaTetenov2018}, \cite{UeharaKY20}, and \cite{Kallus2021}, amongst others, for policy learning under a known shift in characteristics. An analogous weighting could be applied to the welfare gain criterion. These reweighted criteria are typically motivated under the assumption that the CATE is the same under $P$ and $Q$, in which case (\ref{eq:social.welfare.reweight}) induces the same ranking of policies as social welfare in population $Q$. 

To study the implications of shifts in potential outcomes in this setting, consider
\begin{equation}\label{eq:robust.welfare.reweight}
 \inf_{Q \in \mc Q_\rho} \mathrm{W}_\rho(\tau;Q)
\end{equation}
where $\mc Q_\rho = \{ Q : d_{W,\rho}(P,Q) \leq \varepsilon\}$ with
\[
 d_{W,\rho}(P,Q) = \inf_{\pi \in \Pi(P,Q)} \E_\pi[d(Z,\tilde Z)\rho(X)]
\]
for $Z = (X,Y_0,Y_1)$, and where $d$ is the metric (\ref{eq:metric.po}). Suppose that $\rho$ is strictly positive. Criterion (\ref{eq:robust.welfare.reweight}) defines robust welfare over distributions $Q$ in which the marginals for $Y_0$ and $Y_1$ are similar to those under $P$, while $X$ has marginal density $p$; reweighting by $\rho$ converts this to $q$. An identical argument to Proposition~\ref{prop:po.unbounded} yields
\[
 \inf_{Q \in \mc Q_\rho} \mathrm{W}_\rho(\tau; Q) 
 = 
 \mathrm{W}_\rho(\tau; P) - \varepsilon . 
\]
The implications in Remark~\ref{rmk:prop.po} carry over to this criterion. In particular, any policy that maximizes the re-weighted social welfare criterion (\ref{eq:social.welfare.reweight}) must also maximize its robust counterpart (\ref{eq:robust.welfare.reweight}). An analogous result holds for welfare gain.

\section{Proofs of Additional Results}\label{ax:proofs.2}

\subsection{Proofs for Appendix~\ref{ax:bounded}}

\begin{proof}[Proof of Proposition~\ref{prop:ate}]
As $X$ does not appear in the objective, it is without loss of generality to set $b = 0$ and let $P$ and $Q$ be distributions over $(Y_0,Y_1)$. For brevity we just prove the lower bound. The Lagrangian is
\[
 L = \inf_{Q} \sup_{\eta \geq 0 } \E_Q[ Y_1 - Y_0]  + \eta (d_W(P,Q) - \varepsilon) \,.
\]
The Lagrangian dual is
\begin{align*}
 L^\star & = \sup_{\eta \geq 0}  \inf_Q \left( \E_Q[ Y_1  - Y_0] + \eta (d_W(P, Q) - \varepsilon) \right) \\
 & = \sup_{\eta \geq 0} \;\; \inf_Q \inf_{\pi \in \Pi(P,Q)} \E_\pi \bigg[ \tilde Y_1 - \tilde Y_0 + \eta \left(  |Y_0 - \tilde Y_0| + |Y_1 - \tilde Y_1| \right) \bigg] - \eta \varepsilon.
\end{align*} 
Note the iterated infimum over $Q$ and couplings $\pi \in \Pi(P,Q)$ is equivalent to infimizing over all joint distributions for $(Y_0,Y_1,\tilde Y_0,\tilde Y_1)$ with marginal $P$ for $(Y_0,Y_1)$. Hence,
\[
 L^\star = \sup_{\eta \geq 0} \inf_{\{F_{(\tilde Y_0,\tilde Y_1)|(Y_0,Y_1)}\}}  \E_P \bigg[ \E_{F_{(\tilde Y_0,\tilde Y_1)|(Y_0,Y_1)}} \bigg[  \tilde Y_1 -\tilde Y_0  + \eta \left( |Y_0 - \tilde Y_0| + |Y_1 - \tilde Y_1| \right) \bigg| Y_0,Y_1 \bigg] \bigg]  - \eta \varepsilon ,
\]
where the inf is over all conditional distributions for $(\tilde Y_0,\tilde Y_1)$ given $(Y_0,Y_1)$, for each $(Y_0,Y_1)$ in the support of $P$. 
As it is without loss of generality to optimize over point masses,\footnote{A subset of all conditional distributions $F_{(\tilde Y_0,\tilde Y_1)|(Y_0,Y_1)}$ are point masses. The infimum over point masses is an upper bound for the infimum over conditional distributions. Suppose there is a distribution that achieves a lower value than the infimum with point masses. The objective is linear in the distribution, so we can find some point $(\tilde y_0,\tilde y_1)$ in the support of the distribution for which the objective is lower than said infimum, and put a point mass there. This gives a contradiction.} 
\[
 L^\star = \sup_{\eta \geq 0 }\E_P \left[ \inf_{\tilde y_0, \tilde y_1} \left( \tilde y_1 - \tilde y_0 + \eta \left( |Y_0 - \tilde y_0| + |Y_1 - \tilde y_1|  \right) \right) \right] - \eta \varepsilon .
\] 

For the inner problem, first suppose $\eta \in [0,1)$. Note $\tilde y_1 + \eta |Y_1 - \tilde y_1|$ is minimized by taking $\tilde y_1 \downarrow \underline Y$. Similarly, $- \tilde y_0 + \eta |Y_0 - \tilde y_0|$ is minimized by taking $\tilde y_0 \uparrow \overline Y$. Hence,
\[
 \E_P \left[ \inf_{\tilde y_0, \tilde y_1} \left( \tilde y_1 - \tilde y_0 + \eta \left( |Y_0 - \tilde y_0| + |Y_1 - \tilde y_1| \right) \right) \right] 
 = 
 (1 - \eta) \left(  \underline Y - \overline Y \right) + \eta \E_P[Y_1 - Y_0] \, .
\] 
If $\eta \geq 1$, then $\tilde y_1 + \eta |Y_1 - \tilde y_1|$ is minimized by taking $\tilde y_1 = Y_1$ and $- \tilde y_0 + \eta |Y_0 - \tilde y_0|$ is minimized by taking $\tilde y_0 = Y_0$. Hence, 
\[
 \E_P \left[ \inf_{\tilde y_0, \tilde y_1} \left( \tilde y_1 - \tilde y_0 + \eta \left( |Y_0 - \tilde y_0| + |Y_1 - \tilde y_1| \right) \right) \right] = \E_P [Y_1 - Y_0] \,.
\]
Combining the preceding three displays, we obtain $L^\star = \max \left\{ \E_P [Y_1 - Y_0] - \varepsilon \,,\,\underline Y - \overline Y  \right\}$.

Finally, $L = L^\star$ follows by Theorem~1 of \cite{GaoKleywegt}.
\end{proof}

\bigskip

\begin{proof}[Proof of Proposition~\ref{prop:po}]
The Lagrangian and its dual are
\begin{align*}
 L & = \inf_Q \sup_{\eta \geq 0} \left( \E_Q[ Y_1 \tau(X) + Y_0 (1-\tau(X))] + \eta (d_W(P, Q) - \varepsilon) \right) , \\
 L^\star & = \sup_{\eta \geq 0}  \inf_Q \left( \E_Q[ Y_1 \tau(X) + Y_0 (1-\tau(X))] + \eta (d_W(P, Q) - \varepsilon) \right) \\
 & = \sup_{\eta \geq 0} \;\; \inf_Q \inf_{\pi \in \Pi(P,Q)} \E_\pi \bigg[ \tilde Y_1 \tau(\tilde X) + \tilde Y_0 (1-\tau(\tilde X))  \\
 & \quad \quad \quad \quad \quad \quad \quad \quad + \eta \left(  |Y_0 - \tilde Y_0| + |Y_1 - \tilde Y_1|  + \infty \times \mb I[ X \neq \tilde X] \right) \bigg] - \eta \varepsilon.
\end{align*} 
The inf over $Q$ and $\pi \in \Pi(P,Q)$ is equivalent to infimizing over all joint distributions for $(Z,\tilde Z)$ with marginal $P$ for $Z$. It suffices to consider distributions for which $X = \tilde X$ almost surely. Hence,
\begin{multline*}
 L^\star = \sup_{\eta \geq 0} \inf_{\{F_{(\tilde Y_0,\tilde Y_1)|Z}\}}  \E_P \bigg[ \E_{F_{(\tilde Y_0,\tilde Y_1)|Z}} \bigg[  \tilde Y_1 \tau( X) + \tilde Y_0 (1-\tau( X)) \\
  + \eta \left( |Y_0 - \tilde Y_0| + |Y_1 - \tilde Y_1|\right) \bigg| Z \bigg] \bigg] - \eta \varepsilon,
\end{multline*}
where the infimum is over all conditional distributions for $(\tilde Y_0,\tilde Y_1)$ given $Z$, for each $Z$ in the support of $P$. 
As it is without loss of generality to optimize over point masses,
\begin{equation}\label{prop:po.e.1}
 L^\star = \sup_{\eta \geq 0 }\E_P \left[ \inf_{\tilde y_0, \tilde y_1} \left( \tilde y_1 \tau(X) + \tilde y_0 (1-\tau(X)) + \eta \left( |Y_0 - \tilde y_0| + |Y_1 - \tilde y_1| \right) \right) \right] - \eta \varepsilon.
\end{equation}
The remainder of the proof will differ depending on whether $\underline Y > -\infty$ or $\underline Y = -\infty$. 

\medskip

\underline{Case 1: $\underline Y > -\infty$.} The inner infimization with respect to $(\tilde y_0,\tilde y_1)$ in (\ref{prop:po.e.1}) may be solved in closed form for each fixed $Z = (X,Y_0,Y_1)$. Suppose $\eta \in [0,1)$. Then $y \mapsto y + \eta|Y - y|$ is minimized by letting $y \downarrow \underline Y$, with the minimizing value being $\underline Y + \eta (Y - \underline Y)$. 
For $\eta \in [0,1)$, the objective in (\ref{prop:po.e.1}) becomes
\[
  (1-\eta) \underline Y +  \eta \left( \E_P \left[ Y_1 \tau(X) + Y_0 (1-\tau(X))  \right] -  \varepsilon \right).
\]
Maximizing with respect to $\eta \in [0,1)$ yields
\[
 \max \Big\{ 
 \E_P \left[ Y_1 \tau(X) + Y_0 (1-\tau(X)) \right]  - \varepsilon \,,\,
 \underline Y \Big\} \,.
\]
Conversely, if $\eta \geq 1$ then $y + \eta |Y - y|$ is minimized by setting $y = Y$, in which case the objective in (\ref{prop:po.e.1}) is $\E_P \left[ Y_1 \tau(X) + Y_0 (1-\tau(X)) \right]  - \eta \varepsilon$, which is maximized with $\eta = 1$. Combining these results yields $L^\star =\max \{ 
 \E_P \left[ Y_1 \tau(X) + Y_0 (1-\tau(X)) \right]  - \varepsilon \,,\,
 \underline Y \} $.

It remains to show that $L = L^\star$. We provide a constructive proof. By weak duality ($L^\star \leq L$), it suffices to show $L \leq L^\star$. Partition $\mc X = \mc X_0 \cup \mc X_1$ with $\mc X_0 = \tau^{-1}(\{0\})$ and $\mc X_1 = \tau^{-1}(\{1\})$. Let $Q_0$ denote the distribution of $T(Z)$ with $Z \sim P$, where
\[
 T(x,y_0,y_1) = \left\{ \begin{array}{ll}
 (x,\underline Y,y_1) & \mbox{if $x \in \mc X_0$}, \\
 (x,y_0,\underline Y) & \mbox{if $x \in \mc X_1$}. 
 \end{array} \right.
\]
Then $\E_{Q_0}[Y_1 \tau(X) + Y_0 (1-\tau(X)) ] = \underline Y$. Consider the coupling $(Z,T(Z)) \sim \pi$ for $Z \sim P$. Then under the metric (\ref{eq:metric.po}), we have
\begin{align}
 d_W(P,Q_0) & \leq \E_\pi[ d((X,Y_0,Y_1),(\tilde X,\tilde Y_0,\tilde Y_1)) ] \notag \\
 & = \E_P[ (Y_0 - \underline Y)\mb I[X \in \mc X_0] + (Y_1 - \underline Y)\mb I[X \in \mc X_1]] \notag \\
 & = \E_P \left[ Y_1 \tau(X) + Y_0 (1-\tau(X)) \right] - \underline Y. \label{eq:prop:po.1}
\end{align}
It follows that whenever $\E_P \left[ Y_1 \tau(X) + Y_0 (1-\tau(X)) \right] - \varepsilon < \underline Y$, we have
\begin{multline*}
 L \leq \sup_{\eta \geq 0} \left( \E_{Q_0}[ Y_1 \tau(X) + Y_0 (1-\tau(X))] + \eta (d_W(P, Q_0) - \varepsilon) \right) \\
 \leq \underline Y + \sup_{\eta \geq 0} \eta \left( \E_P \left[ Y_1 \tau(X) + Y_0 (1-\tau(X)) \right] - \varepsilon - \underline Y \right)  = L^\star.
\end{multline*}
If $\E_P \left[ Y_1 \tau(X) + Y_0 (1-\tau(X)) \right] - \varepsilon \geq \underline Y$, let $Q_1 = w Q_0 + (1-w)P$ be a mixture with 
\begin{equation}\label{eq:prop:po.2}
 w = \frac{\varepsilon}{\E_P \left[ Y_1 \tau(X) + Y_0 (1-\tau(X)) \right] - \underline Y}\,.
\end{equation}
Then $\E_{Q_1}[Y_1 \tau(X) + Y_0 (1-\tau(X)) ] = \E_P \left[ Y_1 \tau(X) + Y_0 (1-\tau(X)) \right] - \varepsilon$. Moreover, by convexity of $d_W$ (see, e.g., \citeauthor{Villani} (\citeyear{Villani}, Theorem~4.8)),
\[
 d_W(P,Q_1) \leq w d_W(P, Q_0) + (1-w) d_W(P, P) = w d_W(P, Q_0) \leq \varepsilon,
\]
where the final inequality follows from (\ref{eq:prop:po.1}) and (\ref{eq:prop:po.2}). Therefore, we again have
\begin{multline*}
 L \leq \sup_{\eta \geq 0} \left( \E_{Q_1}[ Y_1 \tau(X) + Y_0 (1-\tau(X))] + \eta (d_W(P, Q_1) - \varepsilon) \right) \\
  \leq \E_P \left[ Y_1 \tau(X) + Y_0 (1-\tau(X)) \right] - \varepsilon + \sup_{\eta \geq 0} \eta (d_W(P, Q_1)  - \varepsilon) = L^\star.
\end{multline*}

\medskip

\underline{Case 2: $\underline Y = -\infty$.} Suppose $\eta \in [0,1)$. Then $y + \eta |Y - y|$ is minimized by taking $y \downarrow -\infty$, and the minimizing value is $-\infty$. Conversely, if $\eta \geq 1$ then $y + \eta |Y - y|$ is minimized by setting $y = Y$, so the objective in (\ref{prop:po.e.1}) is $\E_P [  Y_0 + (Y_1 - Y_0)\tau(X)] - \eta \varepsilon$,
which is maximized over $\eta \geq 1$ at $\eta = 1$. Hence,  $L^\star = \E_P \left[ Y_1 \tau(X) + Y_0 (1-\tau(X)) \right]  - \varepsilon$.

It remains to show $L \leq L^\star$. As $\underline Y = -\infty$ and $Y_0$ and $Y_1$ have finite first moments under $P$, there exists $y^* \in \mc Y$ such that
\[
 \E_P \left[ (Y_1 - y^*)_+ \tau(X) + (Y_0 - y^*)_+ (1-\tau(X)) \right] \geq \varepsilon ,
\]
where $(a)_+ = \max\{a,0\}$. Let $Q_0$ denote the distribution of $T(Z)$ with $Z \sim P$, where
\[
 T(x,y_0,y_1) = \left\{ \begin{array}{ll}
 (x,\min\{y_0,y^*\},y_1) & \mbox{if $x \in \mc X_0$}, \\
 (x,y_0,\min\{y_1,y^*\}) & \mbox{if $x \in \mc X_1$}. 
 \end{array} \right.
\]
Then
\[
\begin{aligned}
 \E_{Q_0}[Y_1 \tau(X) + Y_0 (1-\tau(X)) ] & = \E_{P}[\min\{Y_1,y^*\} \tau(X) + \min\{Y_0,y^*\} (1-\tau(X)) ] \\
 & \leq \E_{P}[Y_1 \tau(X) + Y_0 (1-\tau(X)) ] - \varepsilon\,,
\end{aligned}
\]
and $\E_{Q_0}[Y_1 \tau(X) + Y_0 (1-\tau(X)) ]$ is finite because $|\min\{y,y^*\}| \leq |y| + |y^*|$ for all $y$ and $Y_0$ and $Y_1$ have finite first moments under $P$. Moreover, with $(Z,T(Z)) \sim \pi$ for $Z \sim P$, we have
\begin{align}
 d_W(P,Q_0) & \leq \E_\pi[ d((X,Y_0,Y_1),(\tilde X,\tilde Y_0,\tilde Y_1)) ] \notag \\
 & = \E_P[ (Y_0 - y^*)_+ \mb I[X \in \mc X_0] + (Y_1 - y^*)_+ \mb I[X \in \mc X_1]] \notag \\
 & = \E_P \left[ Y_1 \tau(X) + Y_0 (1-\tau(X)) \right] - \E_{Q_0}[Y_1 \tau(X) + Y_0 (1-\tau(X)) ]. \label{eq:prop:po.3}
\end{align}
Let $Q_1 = w Q_0 + (1-w)P$ be a mixture distribution with weight 
\begin{equation}\label{eq:prop:po.4}
 w = \frac{\varepsilon}{\E_P \left[ Y_1 \tau(X) + Y_0 (1-\tau(X)) \right] - \E_{Q_0}[Y_1 \tau(X) + Y_0 (1-\tau(X)) ]}
\end{equation}
on $Q_0$. Then $\E_{Q_1}[Y_1 \tau(X) + Y_0 (1-\tau(X)) ] = \E_P \left[ Y_1 \tau(X) + Y_0 (1-\tau(X)) \right] - \varepsilon$. Moreover,
\[
 d_W(P,Q_1) \leq w d_W(P, Q_0) + (1-w) d_W(P, P) = w d_W(P, Q_0) \leq \varepsilon,
\]
by convexity of $d_W$ and (\ref{eq:prop:po.3}) and (\ref{eq:prop:po.4}). Therefore,
\begin{multline*}
 L \leq \sup_{\eta \geq 0} \left( \E_{Q_1}[ Y_1 \tau(X) + Y_0 (1-\tau(X))] + \eta (d_W(P, Q_1) - \varepsilon) \right) \\
  \leq \E_P \left[ Y_1 \tau(X) + Y_0 (1-\tau(X)) \right] - \varepsilon + \sup_{\eta \geq 0} \eta (d_W(P, Q_1)  - \varepsilon) = L^\star,
\end{multline*}
as required.
\end{proof}

\bigskip

\begin{proof}[Proof of Proposition~\ref{prop:po.gain}]
The proof is similar to the proof of Proposition~\ref{prop:po}.
\end{proof}

\bigskip

\begin{proof}[Proof of Proposition~\ref{prop:po.x}]
We argue as for Proposition~\ref{prop:po}. The Lagrangian is
\[
 L = \inf_Q \sup_{\eta \geq 0} \left( \E_Q[ Y_1 \tau(X) + Y_0 (1-\tau(X))] + \eta (d_W(P, Q) - \varepsilon) \right) 
\]
and its dual is
\begin{multline} \label{eq:po.x.1}
 L^\star  
 = \sup_{\eta \geq 0}  \E_P \bigg[ \inf_{(\tilde x, \tilde y_0, \tilde y_1)} \bigg( \tilde y_1 \tau(\tilde x) + \tilde y_0 (1-\tau(\tilde x)) \\
 + \eta \left( |Y_0 - \tilde y_0| + |Y_1 - \tilde y_1| + \|X - \tilde x\| \right) \bigg) \bigg] - \eta \varepsilon .
\end{multline}

Consider the inner infimization at any fixed $Z = (X,Y_0,Y_1)$. We first fix $\tilde x$ and optimize with respect to $(\tilde y_0,\tilde y_1)$, then optimize with respect to $\tilde x$. There are two cases.

\medskip

\underline{Case 1: $\underline Y > -\infty$.} If $\eta \in [0,1)$, then the infimum with respect to $(\tilde y_0,\tilde y_1)$ (at fixed $\tilde x$) is attained with $\tilde y_0 = Y_0$ and letting $\tilde y_1 \downarrow \underline Y$ when $\tau(\tilde x) = 1$, and letting $\tilde y_0 \downarrow \underline Y$ and $\tilde y_1 = Y_1$ when $\tau(\tilde x) = 0$. For $\eta \in [0,1)$, the objective in (\ref{eq:po.x.1}) becomes
\[
 (1 - \eta) \underline Y + \eta \E_P \left[ Y_0  + \inf_{\tilde x} \left(  (Y_1 - Y_0) \tau(\tilde x) +  \|X - \tilde x\|  \right) \right] - \eta \varepsilon .
\]
Maximizing with respect to $\eta \in [0,1)$ yields
\[
 \max \left\{ \E_P \left[ Y_0  +  \inf_{\tilde x} \left( (Y_1 - Y_0) \tau(\tilde x) +  \|X - \tilde x\|  \right) \right] - \varepsilon , \underline Y \right\}.
\]
If $\eta \geq 1$, then the infimum with respect to $(\tilde y_0,\tilde y_1)$ (at fixed $\tilde x$) is attained with $(\tilde y_0,\tilde y_1) = (Y_0,Y_1)$ and the objective in (\ref{eq:po.x.1}) becomes
\begin{equation} \label{eq:prop.po.x.1}
  \E_P \left[ Y_0 + \inf_{\tilde x} \left( (Y_1 - Y_0)\tau(\tilde x) +  \eta \left(\|X - \tilde x\| \right)\right) \right]  - \eta \varepsilon.
\end{equation}

\medskip

\underline{Case 2: $\underline Y = -\infty$.} If $\eta \in [0,1)$, then the infimum is $-\infty$, which is obtained by taking $\tilde y_0 \downarrow -\infty$ if $\tau(\tilde x) = 0$ and $\tilde y_1 \downarrow -\infty$ if $\tau(\tilde x) = 1$. Conversely if $\eta \geq 1$ then the infimum is attained with $(\tilde y_0,\tilde y_1) = (Y_0,Y_1)$ and the dual objective again reduces to (\ref{eq:prop.po.x.1}). 

Combining these results, we obtain
\[
 L^\star = \max \left\{ \sup_{\eta \geq 1} \E_P \left[ Y_0  +  \inf_{\tilde x} \left( (Y_1 - Y_0) \tau(\tilde x) + \eta \|X - \tilde x\|  \right) \right] - \eta \varepsilon \,,\, \underline Y \right\}  .
\]
We may split the infimization up into separate infimizations over $\{\tilde x : \tau(\tilde x) = 0\}$ and $\{\tilde x : \tau(\tilde x) = 1\}$, then take the minimum:
\begin{align*}
 L^\star & =  \max \left\{ \sup_{\eta \geq 1} \E_P \left[ Y_0  +   \min \left\{ Y_1 - Y_0 +  \inf_{\tilde x:\tau(\tilde x) = 1}  \eta \|X - \tilde x\|  , \inf_{\tilde x:\tau(\tilde x) = 0}  \eta \|X - \tilde x\| \right\} \right] - \eta \varepsilon \,,\, \underline Y \right\} \\
 & =  \max \left\{ \sup_{\eta \geq 1} \E_P \left[ Y_0 + \min \left\{   Y_1  - Y_0 + \eta h_1(X; \tau) , \eta h_0(X; \tau) \right\} \right] - \eta \varepsilon \,,\, \underline Y \right\} \,.
\end{align*}

Finally, strong duality holds by Theorem~1 of \cite{GaoKleywegt}.
\end{proof}

\bigskip

\begin{proof}[Proof of Proposition~\ref{prop:po.gain.x}]
First note that if $\tau \equiv 0$ then $\mr{WG}(\tau;Q) = 0$ for all $Q$, so $\mr{RWG}(\tau;P) = \mr{WG}(\tau;P) = 0$. Now suppose that $\tau \not \equiv 0$. 
We argue as in the proof of Proposition~\ref{prop:po.x}, stating only the necessary modifications. The dual is
\begin{multline} \label{eq:po.x.gain.1}
 L^\star  
 = \sup_{\eta \geq 0}  \E_P \bigg[ \inf_{(\tilde x, \tilde y_0, \tilde y_1)} \bigg( (\tilde y_1 - \tilde y_0) \tau(\tilde x) + \eta \left( |Y_0 - \tilde y_0| + |Y_1 - \tilde y_1| + \|X - \tilde x\| \right) \bigg) \bigg] - \eta \varepsilon .
\end{multline}

\underline{Case 1: $-\infty < \underline Y \leq \overline Y < \infty$.} If $\eta \in [0,1)$, then the infimum with respect to $(\tilde y_0,\tilde y_1)$ (at fixed $\tilde x$) is attained by letting $\tilde y_0 \uparrow \overline Y$ and $\tilde y_1 \downarrow \underline Y$ if $\tau(\tilde x) = 1$ and $(\tilde y_0, \tilde y_1) = (Y_0,Y_1)$ if $\tau(\tilde x) = 0$. For $\eta \in [0,1)$, the expectation term in (\ref{eq:po.x.gain.1}) becomes
\begin{multline*}
 \E_P \left[ \inf_{\tilde x} \left(  \tau(\tilde x) (\ul Y - \ol Y + \eta(\ol Y - Y_0) + \eta(Y_1 - \ul Y)) + \eta \|X - \tilde x\| \right) \right] \\
 = \E_P \left[ \min\left\{ (1-\eta)(\ul Y - \ol Y) + \eta(Y_1 - Y_0) + \eta h_1(X;\tau) , \eta h_0(X;\tau) \right\} \right] \,.
\end{multline*}
If $\eta \geq 1$ then the infimum is achieved with $(\tilde y_0,\tilde y_1) = (Y_0,Y_1)$ and the expectation term in (\ref{eq:po.x.gain.1}) becomes
\begin{multline}
 \E_P \left[ \inf_{\tilde x} \left(  \tau(\tilde x) (Y_1 - Y_0) + \eta \|X - \tilde x\| \right) \right] \\
 = \E_P \left[ \min\left\{ (Y_1 - Y_0) + \eta h_1(X;\tau) , \eta h_0(X;\tau) \right\} \right] \,. \label{eq:po.x.gain.2}
\end{multline}

\medskip

\underline{Case 2: $\underline Y = -\infty$ and/or $\overline Y = +\infty$.} If $\eta \in [0,1)$, then the infimum is $-\infty$, which is obtained by letting $\tilde y_0 \uparrow +\infty$ if $\ol Y = +\infty$ and/or $\tilde y_1 \downarrow -\infty$ if $\ul Y = -\infty$ for any $\tilde x$ for which $\tau(\tilde x) = 1$ (and at least one such $\tilde x$ exists because $\tau \not \equiv 0$). Conversely if $\eta \geq 1$ then the infimum is achieved with $(\tilde y_0, \tilde y_1) = (Y_0,Y_1)$ and the expectation reduces to~(\ref{eq:po.x.gain.2}). 

Strong duality again holds by Theorem~1 of \cite{GaoKleywegt}.
\end{proof}

\subsection{Proofs for Appendix~\ref{ax:extensions}}

%
%
%
%
%

\begin{proof}[Proof of Proposition~\ref{prop:po.Wp}]
We argue as in the proof of Proposition~\ref{prop:po}. It suffices to consider $p > 1$ as $p = 1$ was proved already for Proposition~\ref{prop:po}. We have
\begin{align*}
 L & = \inf_Q \sup_{\eta \geq 0} \left( \E_Q[ Y_1 \tau(X) + Y_0 (1-\tau(X))] + \frac{\eta}{p} (d_{W,p} ( P, Q)^p - \varepsilon^p) \right) , \\
 L^\star & = \sup_{\eta \geq 0 }\E_P \left[ \inf_{\tilde y_0, \tilde y_1} \left( \tilde y_1 \tau(X) + \tilde y_0 (1-\tau(X)) + \frac{\eta}{p} \left( |Y_0 - \tilde y_0|^p + |Y_1 - \tilde y_1|^p - \varepsilon^p \right) \right) \right] .
\end{align*}
When $\eta = 0$ the inner infimum is $-\infty$. Suppose $\eta > 0$. Fix $Z = (X,Y_0,Y_1)$. If $\tau(X) = 1$, then the inner infimum over $\tilde y_0$ is attained at $\tilde y_0 = Y_0$. The inner infimum over $\tilde y_1$ reduces to minimizing $y \mapsto y + \frac{\eta}{p}|Y_1-y|^p$, which is achieved at $y = Y_1 - \eta^{-1/(p-1)}$. A symmetric argument applies when $\tau(X) = 0$. Hence,
\[
 \begin{aligned}
 L^{\star} & = \sup_{\eta > 0} \ \mathbb{E}_{P}[Y_1 \tau(X) + Y_0 (1 - \tau(X))] - \eta^{-1/(p-1)} + \frac{\eta}{p} \eta^{-p/(p-1)} - \frac{\eta}{p}\varepsilon^p \\
 & = \E_{P}[Y_1 \tau(X) + Y_0 (1 - \tau(X))] - \varepsilon.
\end{aligned}
\]
By weak duality we have $L^\star \leq L$. It remains to show $L \leq L^\star$. To this end, let $Q$ denote the distribution of $T(Z)$, where $T$ is given in~\eqref{eq:t.map}. Then $d_{W,p}(P, Q) \leq \varepsilon$ and $\mathrm{W}(\tau;Q) = \mathrm{W}(\tau;P) - \varepsilon$, so $L \leq L^\star$.
\end{proof}

{

\let\oldbibliography\thebibliography
\renewcommand{\thebibliography}[1]{\oldbibliography{#1}
\setlength{\itemsep}{0pt}}

\putbib
}
\end{bibunit}


\begin{thebibliography}{}

\bibitem[\protect\citeauthoryear{Abadie}{Abadie}{2003}]{Abadie2003}
Abadie, A. (2003).
\newblock Semiparametric instrumental variable estimation of treatment response
  models.
\newblock {\em Journal of Econometrics\/}~{\em 113\/}(2), 231--263.

\bibitem[\protect\citeauthoryear{Abbring and Heckman}{Abbring and
  Heckman}{2007}]{AbbringHeckman2007}
Abbring, J.~H. and J.~J. Heckman (2007).
\newblock Econometric evaluation of social programs, {Part III}:
  {D}istributional treatment effects, dynamic treatment effects, dynamic
  discrete choice, and general equilibrium policy evaluation.
\newblock Volume~6 of {\em Handbook of Econometrics}, Chapter~72, pp.\
  5145--5303. Elsevier.

\bibitem[\protect\citeauthoryear{Allcott}{Allcott}{2015}]{Allcott2015}
Allcott, H. (2015).
\newblock Site selection bias in program evaluation.
\newblock {\em The Quarterly Journal of Economics\/}~{\em 130\/}(3),
  1117--1165.

\bibitem[\protect\citeauthoryear{Athey and Wager}{Athey and
  Wager}{2021}]{AtheyWager2021}
Athey, S. and S.~Wager (2021).
\newblock Policy learning with observational data.
\newblock {\em Econometrica\/}~{\em 89\/}(1), 133--161.

\bibitem[\protect\citeauthoryear{Banerjee and Duflo}{Banerjee and
  Duflo}{2009}]{BannerjeeDuflo2009}
Banerjee, A.~V. and E.~Duflo (2009).
\newblock The experimental approach to development economics.
\newblock {\em Annual Review of Economics\/}~{\em 1}, 151--178.

\bibitem[\protect\citeauthoryear{Bartlett, Jordan, and McAuliffe}{Bartlett
  et~al.}{2006}]{BJM2006}
Bartlett, P.~L., M.~I. Jordan, and J.~D. McAuliffe (2006).
\newblock Convexity, classification, and risk bounds.
\newblock {\em Journal of the American Statistical Association\/}~{\em
  101\/}(473), 138--156.

\bibitem[\protect\citeauthoryear{Ben-Michael, Greiner, Imai, and
  Jiang}{Ben-Michael
  et~al.}{2025}]{benmichael2025safepolicylearningextrapolation}
Ben-Michael, E., D.~J. Greiner, K.~Imai, and Z.~Jiang (2025).
\newblock Safe policy learning through extrapolation: Application to pre-trial
  risk assessment.
\newblock {\em Journal of the American Statistical Association\/}~{\em
  120\/}(551), 1386--1399.

\bibitem[\protect\citeauthoryear{Bhattacharya and Dupas}{Bhattacharya and
  Dupas}{2012}]{BhattacharyaDupas}
Bhattacharya, D. and P.~Dupas (2012).
\newblock Inferring welfare maximizing treatment assignment under budget
  constraints.
\newblock {\em Journal of Econometrics\/}~{\em 167\/}(1), 168--196.

\bibitem[\protect\citeauthoryear{D'Adamo}{D'Adamo}{2022}]{dadamo2022orthogonalpolicylearningambiguity}
D'Adamo, R. (2022).
\newblock Orthogonal policy learning under ambiguity.
\newblock {\em arXiv preprint arXiv:2111.10904\/}.

\bibitem[\protect\citeauthoryear{Deaton}{Deaton}{2010}]{Deaton2010}
Deaton, A. (2010).
\newblock Instruments, randomization, and learning about development.
\newblock {\em Journal of Economic Literature\/}~{\em 48\/}(2), 424--455.

\bibitem[\protect\citeauthoryear{Fan and Park}{Fan and
  Park}{2010}]{FanPark2010}
Fan, Y. and S.~S. Park (2010).
\newblock Sharp bounds on the distribution of treatment effects and their
  statistical inference.
\newblock {\em Econometric Theory\/}~{\em 26\/}(3), 931--951.

\bibitem[\protect\citeauthoryear{García and Saavedra}{García and
  Saavedra}{2017}]{GS2017}
García, S. and J.~E. Saavedra (2017).
\newblock Educational impacts and cost-effectiveness of conditional cash
  transfer programs in developing countries: A meta-analysis.
\newblock {\em Review of Educational Research\/}~{\em 87\/}(5), 921--965.

\bibitem[\protect\citeauthoryear{Gechter}{Gechter}{2024}]{Gechter2024}
Gechter, M. (2024).
\newblock Generalizing the results from social experiments: Theory and evidence
  from {I}ndia.
\newblock {\em Journal of Business \& Economic Statistics\/}~{\em 42\/}(2),
  801--811.

\bibitem[\protect\citeauthoryear{Giné and Nickl}{Giné and
  Nickl}{2016}]{GineNickl2016}
Giné, E. and R.~Nickl (2016).
\newblock {\em Mathematical Foundations of Infinite-Dimensional Statistical
  Models}.
\newblock Cambridge Series in Statistical and Probabilistic Mathematics.
  Cambridge University Press.

\bibitem[\protect\citeauthoryear{Heckman, Smith, and Clements}{Heckman
  et~al.}{1997}]{HeckmanSmithClements1997}
Heckman, J.~J., J.~Smith, and N.~Clements (1997).
\newblock {Making The Most Out Of Programme Evaluations and Social Experiments:
  Accounting For Heterogeneity in Programme Impacts}.
\newblock {\em The Review of Economic Studies\/}~{\em 64\/}(4), 487--535.

\bibitem[\protect\citeauthoryear{Imbens and Menzel}{Imbens and
  Menzel}{2021}]{ImbensMenzel2021}
Imbens, G. and K.~Menzel (2021).
\newblock {A causal bootstrap}.
\newblock {\em The Annals of Statistics\/}~{\em 49\/}(3), 1460--1488.

\bibitem[\protect\citeauthoryear{Kallus}{Kallus}{2017}]{Kallus2017}
Kallus, N. (2017).
\newblock Recursive partitioning for personalization using observational data.
\newblock In {\em Proceedings of the 34th International Conference on Machine
  Learning}, Volume~70, pp.\  1789--1798.

\bibitem[\protect\citeauthoryear{Kallus}{Kallus}{2021}]{Kallus2021}
Kallus, N. (2021).
\newblock More efficient policy learning via optimal retargeting.
\newblock {\em Journal of the American Statistical Association\/}~{\em
  116\/}(534), 646--658.

\bibitem[\protect\citeauthoryear{Kallus and Zhou}{Kallus and
  Zhou}{2021}]{KZ2021}
Kallus, N. and A.~Zhou (2021).
\newblock Minimax-optimal policy learning under unobserved confounding.
\newblock {\em Management Science\/}~{\em 67\/}(5), 2870--2890.

\bibitem[\protect\citeauthoryear{Kido}{Kido}{2022}]{Kido2022}
Kido, D. (2022).
\newblock Distributionally robust policy learning with {W}asserstein distance.
\newblock {\em arXiv preprint arXiv:2205.04637\/}.

\bibitem[\protect\citeauthoryear{Kitagawa and Tetenov}{Kitagawa and
  Tetenov}{2018}]{KitagawaTetenov2018}
Kitagawa, T. and A.~Tetenov (2018).
\newblock Who should be treated? {E}mpirical welfare maximization methods for
  treatment choice.
\newblock {\em Econometrica\/}~{\em 86\/}(2), 591--616.

\bibitem[\protect\citeauthoryear{Lei, Sahoo, and Wager}{Lei
  et~al.}{2023}]{LSW2023}
Lei, L., R.~Sahoo, and S.~Wager (2023).
\newblock Policy learning under biased sample selection.
\newblock {\em arXiv preprint arXiv:2304.11735\/}.

\bibitem[\protect\citeauthoryear{Manski}{Manski}{1997}]{Manski1997}
Manski, C.~F. (1997).
\newblock The mixing problem in programme evaluation.
\newblock {\em The Review of Economic Studies\/}~{\em 64\/}(4), 537--553.

\bibitem[\protect\citeauthoryear{Manski}{Manski}{2004}]{Manski2004}
Manski, C.~F. (2004).
\newblock Statistical treatment rules for heterogeneous populations.
\newblock {\em Econometrica\/}~{\em 72\/}(4), 1221--1246.

\bibitem[\protect\citeauthoryear{Mbakop and Tabord-Meehan}{Mbakop and
  Tabord-Meehan}{2021}]{MTM}
Mbakop, E. and M.~Tabord-Meehan (2021).
\newblock Model selection for treatment choice: Penalized welfare maximization.
\newblock {\em Econometrica\/}~{\em 89\/}(2), 825--848.

\bibitem[\protect\citeauthoryear{McKenzie and Puerto}{McKenzie and
  Puerto}{2021}]{MKP2021}
McKenzie, D. and S.~Puerto (2021).
\newblock Growing markets through business training for female entrepreneurs: A
  market-level randomized experiment in {K}enya.
\newblock {\em American Economic Journal: Applied Economics\/}~{\em 13\/}(2),
  297--332.

\bibitem[\protect\citeauthoryear{Milgrom and Segal}{Milgrom and
  Segal}{2002}]{MilgromSegal2002}
Milgrom, P. and I.~Segal (2002).
\newblock Envelope theorems for arbitrary choice sets.
\newblock {\em Econometrica\/}~{\em 70\/}(2), 583--601.

\bibitem[\protect\citeauthoryear{Mo, Qi, and Liu}{Mo et~al.}{2021}]{MoQiLiu}
Mo, W., Z.~Qi, and Y.~Liu (2021).
\newblock Learning optimal distributionally robust individualized treatment
  rules.
\newblock {\em Journal of the American Statistical Association\/}~{\em
  116\/}(534), 659--674.

\bibitem[\protect\citeauthoryear{Munro}{Munro}{2025}]{Munro2023}
Munro, E. (2025).
\newblock Treatment allocation with strategic agents.
\newblock {\em Management Science\/}~{\em 71\/}(1), 123--145.

\bibitem[\protect\citeauthoryear{Qi, Pang, and Liu}{Qi
  et~al.}{2023}]{QiPangLiu2022}
Qi, Z., J.-S. Pang, and Y.~Liu (2023).
\newblock On robustness of individualized decision rules.
\newblock {\em Journal of the American Statistical Association\/}~{\em
  118\/}(543), 2143--2157.

\bibitem[\protect\citeauthoryear{Qian and Murphy}{Qian and
  Murphy}{2011}]{QianMurphy2011}
Qian, M. and S.~A. Murphy (2011).
\newblock {Performance guarantees for individualized treatment rules}.
\newblock {\em The Annals of Statistics\/}~{\em 39\/}(2), 1180--1210.

\bibitem[\protect\citeauthoryear{Si, Zhang, Zhou, and Blanchet}{Si
  et~al.}{2020}]{SZZB}
Si, N., F.~Zhang, Z.~Zhou, and J.~Blanchet (2020).
\newblock Distributionally robust policy evaluation and learning in offline
  contextual bandits.
\newblock In {\em Proceedings of the 37th International Conference on Machine
  Learning}, Volume 119, pp.\  8884--8894.

\bibitem[\protect\citeauthoryear{Spini}{Spini}{2021}]{Spini2021}
Spini, P.~E. (2021).
\newblock Robustness, heterogeneous treatment effects and covariate shifts.
\newblock {\em arXiv preprint arXiv:2112.09259\/}.

\bibitem[\protect\citeauthoryear{Stoye}{Stoye}{2010}]{Stoye2010}
Stoye, J. (2010).
\newblock Partial identification of spread parameters.
\newblock {\em Quantitative Economics\/}~{\em 1\/}(2), 323--357.

\bibitem[\protect\citeauthoryear{Stoye}{Stoye}{2012}]{Stoye2012}
Stoye, J. (2012).
\newblock Minimax regret treatment choice with covariates or with limited
  validity of experiments.
\newblock {\em Journal of Econometrics\/}~{\em 166\/}(1), 138--156.

\bibitem[\protect\citeauthoryear{Swaminathan and Joachims}{Swaminathan and
  Joachims}{2015}]{SwaminathanJoachims}
Swaminathan, A. and T.~Joachims (2015).
\newblock Batch learning from logged bandit feedback through counterfactual
  risk minimization.
\newblock {\em Journal of Machine Learning Research\/}~{\em 16\/}(52),
  1731--1755.

\bibitem[\protect\citeauthoryear{van~der Vaart and Wellner}{van~der Vaart and
  Wellner}{1996}]{vanderVaartWellner}
van~der Vaart, A.~W. and J.~A. Wellner (1996).
\newblock {\em Weak convergence and empirical processes}.
\newblock Springer.

\bibitem[\protect\citeauthoryear{Vuong and Xu}{Vuong and
  Xu}{2017}]{VuongXu2017}
Vuong, Q. and H.~Xu (2017).
\newblock Counterfactual mapping and individual treatment effects in
  nonseparable models with binary endogeneity.
\newblock {\em Quantitative Economics\/}~{\em 8\/}(2), 589--610.

\bibitem[\protect\citeauthoryear{Zhao, Zeng, Rush, and Kosorok}{Zhao
  et~al.}{2012}]{ZZRK}
Zhao, Y., D.~Zeng, A.~J. Rush, and M.~R. Kosorok (2012).
\newblock Estimating individualized treatment rules using outcome weighted
  learning.
\newblock {\em Journal of the American Statistical Association\/}~{\em
  107\/}(499), 1106--1118.

\end{thebibliography}


\begin{thebibliography}{}

\bibitem[\protect\citeauthoryear{Gao and Kleywegt}{Gao and
  Kleywegt}{2023}]{GaoKleywegt}
Gao, R. and A.~Kleywegt (2023).
\newblock {Distributionally Robust Stochastic Optimization with Wasserstein
  Distance}.
\newblock {\em Mathematics of Operations Research\/}~{\em 48\/}(2), 603--655.

\bibitem[\protect\citeauthoryear{Kallus}{Kallus}{2021}]{Kallus2021}
Kallus, N. (2021).
\newblock More efficient policy learning via optimal retargeting.
\newblock {\em Journal of the American Statistical Association\/}~{\em
  116\/}(534), 646--658.

\bibitem[\protect\citeauthoryear{Kitagawa and Tetenov}{Kitagawa and
  Tetenov}{2018}]{KitagawaTetenov2018}
Kitagawa, T. and A.~Tetenov (2018).
\newblock Who should be treated? {E}mpirical welfare maximization methods for
  treatment choice.
\newblock {\em Econometrica\/}~{\em 86\/}(2), 591--616.

\bibitem[\protect\citeauthoryear{Uehara, Kato, and Yasui}{Uehara
  et~al.}{2020}]{UeharaKY20}
Uehara, M., M.~Kato, and S.~Yasui (2020).
\newblock Off-policy evaluation and learning for external validity under a
  covariate shift.
\newblock In {\em Advances in Neural Information Processing Systems 33}.

\bibitem[\protect\citeauthoryear{Villani}{Villani}{2009}]{Villani}
Villani, C. (2009).
\newblock {\em Optimal Transport: Old and New}.
\newblock Springer.

\end{thebibliography}
\end{document}